\documentclass{aa}  

\usepackage{graphicx}
\usepackage{txfonts}
\usepackage{hyperref}
\usepackage[normalem]{ulem}
\usepackage{cancel}
\usepackage{amsmath}
\defcitealias{Laibe2014a}{LP14a}
\defcitealias{Laibe2014b}{LP14b}

\begin{document}
\title{Modeling dust dynamics in \texttt{OpenGadget3}}
\subtitle{I. SPH implementation of the One-Fluid model}
\author{G. Tedeschi-Prades \inst{1}\thanks{\scriptsize Corresponding author: \email{giovanni.tedeschi@campus.lmu.de}}
\and 
T. Birnstiel\inst{1,2}\fnmsep 
\and 
K. Dolag\inst{1,3}\fnmsep 
\and B. Ercolano\inst{1,2}\fnmsep 
\and M. Hutchison\inst{1,4}}

\institute{University Observatory, Faculty of Physics, Ludwig-Maximilians-Universität München, Scheinerstr. 1, 81679 Munich, Germany 
\and 
Exzellenzcluster ORIGINS, Boltzmannstr. 2, D-85748 Garching, Germany 
\and
Max-Planck-Institut für Astrophysik, Karl-Schwarzschild-Straße 1, 85741 Garching, Germany 
\and 
Hochschule für angewandte Wissenschaften München, Lothstraße 34, D-80335 München, Germany}
\date{Received XXX; accepted YYY}

\abstract
{Dust dynamics plays a critical role in astrophysical processes and has been modeled in hydrodynamical simulations using various approaches. Among particle-based methods like Smoothed Particle Hydrodynamics (SPH), the One-Fluid model has proven to be highly effective for simulating gas-dust mixtures.}
{This study presents the implementation of the One-Fluid model in \texttt{OpenGadget3}, introducing improvements to the original formulation. These enhancements include time-dependent artificial viscosity and conductivity, as well as a novel treatment of dust diffusion using a pressure-like term.}
{The improved model is tested using a suite of dust dynamics benchmark problems: DUSTYBOX, DUSTYWAVE, and DUSTYSHOCK, with the latter extended to multidimensional scenarios, as well as a dusty Sedov-Taylor blast wave. Additional tests include simulations of Cold Keplerian Disks, dusty protoplanetary disks, and Kelvin-Helmholtz instabilities to evaluate the model's robustness in more complex flows.}
{The implementation successfully passes all standard benchmark tests. It demonstrates stability and accuracy in both simple and complex simulations. The new diffusion term improves the handling of flows with large dust-to-gas ratios and low drag coefficients, although limitations of the One-Fluid model in these regimes remain.}
{The enhanced One-Fluid model is a reliable and robust tool for simulating dust dynamics in \texttt{OpenGadget3}. While it retains some limitations inherent to the original formulation, the introduced improvements expand its applicability and address some challenges in gas-dust dynamics.}
\maketitle
\section{Introduction}
Dust, alongside gas, is one of the fundamental components of most astrophysical systems. From the cold interstellar medium to molecular clouds and protoplanetary disks, dust is observed ubiquitously across the Milky Way and other galaxies. Importantly, in many cases, dust is the only directly observable component, detected through its millimeter and infrared thermal emission. Gas properties are often inferred indirectly from dust observations, typically assuming a constant dust-to-gas ratio of $10^{-2}$. Dust also plays a critical role in a variety of astrophysical processes. It contributes to the cooling of gas in star-forming regions \citep{Vogelsberger2019}, serves as a surface for chemical reactions that form complex molecules \citep{Minissale2016}, and provides the primary opacity source in many environments, impacting radiative transfer. Understanding the behavior of dust is therefore crucial not only for interpreting observations but also for modeling the physical conditions and evolution of astrophysical systems.\\
While dust is generally assumed to be perfectly coupled to the gas, this assumption breaks down either for larger dust-to-gas ratios or for larger grain sizes. Protoplanetary disks are the astrophysical environment where dust dynamics has been most thoroughly studied. Here, the uncoupling of dust and gas is essential to understand the evolution of the disk and the early stages of planet formation \citep{Birnstiel2024}. The radial drift of dust grains inward, one of the most important dynamical mechanisms to understand the evolution of disks, is caused by a slight difference between the rotational velocity of gas and of dust.\\
But disks are not the only objects in which the dynamics of dust grains play a role. The outflows of AGB stars are also believed to be accelerated by radiation pressure on dust grains, which consequently accelerates gas by aerodynamic drag. The same mechanism is also proposed to fuel outflows from "dusty torus" regions around AGNs \citep{Soliman2023}.\\
Given the importance of dust as a key component of many astrophysical systems, the inclusion of dust dynamics in modern hydrodynamical simulation codes has seen significant growth in recent years. One widely used approach involves post-processing gas-only simulations to study the behavior of dust. While this method improves upon the assumption of perfect coupling between dust and gas, it neglects the critical back-reaction of dust on gas motion. This omission can lead to inaccuracies, especially in scenarios with substantial dust-to-gas ratios or large grain sizes, where dust can significantly influence the gas dynamics. Self-consistent models of dust dynamics have been implemented in Eulerian codes, both as tracer particles, such as in \texttt{AREPO} \citep{McKinnon2018}, or as additional fluids, like in \texttt{ATHENA++} \citep{Huang2022} and \texttt{FARGO3D} \citep{BenitezLlabay2019}. In Smoothed Particle Hydrodynamics (SPH) frameworks, the model that has achieved the most success in simulating dust dynamics is the One-Fluid model developed by \citep{Laibe2014a}. This approach represents a significant departure from traditional multi-fluid methods by describing the motion of a gas-dust mixture self-consistently, rather than treating gas and dust as separate fluids coupled via aerodynamic drag. A notable implementation of the One-Fluid model, in the terminal velocity approximation limit, is in the SPH code \texttt{PHANTOM} \citep{Price2018}. While this model has been largely successful at simulating the dynamics of small dust grains, it still presents challenges when simulating large dust grains. Most notably, \citep{Laibe2014b} observed that interpenetrating dust flows of weakly coupled grains can violate the assumptions underlying the fluid approximation for a pressureless dust model, causing the code to crash. More subtly, the inability of a pressureless fluid to capture the diffusive nature of a multi-valued velocity distribution leads to artificial clumping of dust, which can produce numerical failures or result in an overestimate of planetesimal formation timescales. This has led to the One-Fluid model to be used mostly in the terminal velocity approximation, where dust is assumed to couple efficiently to the gas on timescales smaller than the numerical integration timestep. One possible solution to this long-standing issue of simulating dust dynamics is the introduction of some form of dust pressure. For instance, \cite{Lynch2024} found that a treatment of dust pressure is needed in order to correctly simulate dust dynamics in debris disks.\\
In this paper, we will describe the implementation of the full (i.e., not in the terminal velocity approximation) One-Fluid model for dust dynamics in \texttt{OpenGadget3}. In particular, we will focus on improvements on the original implementation derived in \citep{Laibe2014b}, such as time-dependent artificial viscosity and conductivity, and a model of dust diffusion via a dust pressure-like term to treat clumping in simulations of weakly coupled dust. While the limitations of the full One-Fluid model are still present in the most extreme cases, we show how these improvements can make the One-Fluid model suitable for a larger set of astrophysical applications compared to the terminal velocity approximation.\\
The paper is organized as follows. In Section 2 we show how the One-Fluid model is derived from the equations describing the motion of dust and gas separately and we show the discretization of the continuum equations in the SPH framework, including the improved viscosity and conductivity terms and the dust-pressure diffusion model. In Section 3, we show the performance of the code on a suite of standard numerical tests, as well as on more complex simulations such as the Cold Keplerian Disk and the Kelvin-Helmholtz instability. Finally, in Section 4, we present our conclusions and outline potential avenues for future development.
\section{Methods}
\subsection{The One-Fluid model}
\label{sec:2.1}
Our implementation of dust dynamics relies and improves on the One-Fluid model introduced in \cite{Laibe2014a}. The equations governing gas dynamics in the presence of dust closely resemble the standard Euler equations, with additional terms in the momentum and energy conservation equations to account for the drag exerted by dust on the gas
\begin{equation}
    \label{eq:continuty_gas}
    \partial_t\rho_g + \nabla\cdot(\rho_g\vec{v}_g) = 0,
\end{equation}
\begin{equation}
    \label{eq:momentum_gas}
    \partial_t \vec{v}_g + (\vec{v}_g\cdot\nabla)\vec{v}_g = -\frac{\nabla P}{\rho_g} + \frac{K}{\rho_g}(\vec{v}_d-\vec{v}_g) + \vec{f},
\end{equation}
\begin{equation}
    \label{eq:energy_gas}
    \partial_t u + (\vec{v}_g\cdot\nabla)u = -\frac{P}{\rho_g}(\nabla\cdot\vec{v}_g) + \frac{K}{\rho_g}(\vec{v}_d-\vec{v}_g)^2 + \Lambda,
\end{equation}
where $u$ and $P$ are the internal energy and the pressure of the gas, respectively, and $K$ is the drag coefficient determining how strongly the gas and the dust are coupled. For simplicity, we will ignore any external force acting on the fluid, $\vec{f}$, as well as any external heating, $\Lambda$.\\
The equations for dust are essentially describing a pressureless fluid, with an additional term in the momentum equation describing the drag of gas onto the dust, similar and opposite to the one in Eq.\ref{eq:momentum_gas}
\begin{equation}
    \label{eq:continutiy_dust}
    \partial_t \rho_d + \nabla\cdot(\rho_d\vec{v}_d)=0,
\end{equation}
\begin{equation}
    \label{eq:momentum_dust}
    \partial_t\vec{v}_d+(\vec{v}_d\cdot\nabla)\vec{v}_d = -\frac{K}{\rho_d}(\vec{v}_d - \vec{v}_g) + \vec{f}.
\end{equation}
As in the case of the gas momentum equation, we will ignore the term $\vec{f}$ as well as any effect of gas pressure on the dust.\\
\cite{Laibe2014a} show that, by introducing a new and convenient set of variables, the equations describing the motion of gas and dust can be combined into a set of equations describing the motion of the mixture of both fluids. In particular, the following variables are used
\begin{equation}
    \rho = \rho_g + \rho_d,
\end{equation}
\begin{equation}
    \epsilon = \frac{\rho_d}{\rho_g + \rho_d},
\end{equation}
\begin{equation}
    \vec{v} = \frac{\rho_g\vec{v}_g + \rho_d\vec{v}_d}{\rho_g + \rho_d},
\end{equation}
\begin{equation}
    \vec{\Delta v} = \vec{v_d} - \vec{v_g},
\end{equation}
which described, respectively, the total density ($\rho$), the fraction of dust density compared to the total density ($\epsilon$), the barycenter velocity of the mixture ($\vec{v}$), and the velocity difference between the dust and the gas phases ($\vec{\Delta v}$). By using these variables, the Eq. \ref{eq:continuty_gas} - \ref{eq:momentum_dust} can be rewritten as
\begin{equation}
    \label{eq:continuity}
    \frac{d\rho}{dt} = -\rho(\nabla\cdot\vec{v}),
\end{equation}
\begin{equation}
    \label{eq:dust_frac}
    \frac{d\epsilon}{dt} = -\frac{1}{\rho}\nabla\cdot[\epsilon(1-\epsilon)\rho\vec{\Delta v}],
\end{equation}
\begin{equation}
    \label{eq:momentum}
    \frac{d\vec{v}}{dt} = -\frac{\nabla P}{\rho} - \frac{1}{\rho}\nabla\cdot[\epsilon(1-\epsilon)\rho\vec{\Delta v}\vec{\Delta v}],
\end{equation}
\begin{equation}
    \label{eq:DeltaVel}
    \frac{d\vec{\Delta v}}{dt} =  \frac{\nabla P}{\rho(1-\epsilon)} -\frac{\vec{\Delta v}}{t_s} - (\vec{\Delta v}\cdot\nabla)\vec{v} + \vec{\Delta v}(\vec{\Delta v \cdot \nabla})\epsilon + (2\epsilon -1 )(\vec{\Delta v}\cdot \nabla)\vec{\Delta v},
\end{equation}
\begin{equation}
    \label{eq:energy}
    \frac{du}{dt} = -\frac{P}{\rho(1-\epsilon)}\nabla\cdot(\vec{v}-\epsilon\vec{\Delta v}) + \epsilon(\vec{\Delta v}\cdot \nabla)u + \epsilon\frac{\vec{\Delta v}^2}{t_s},
\end{equation}
where we already introduced the Lagrangian, or co-moving, temporal derivative
\begin{equation}
    \frac{d}{dt} = \frac{\partial}{\partial t} + (\vec{v}\cdot \nabla),
\end{equation}
and the stopping time $t_s$, which is related to the drag coefficient by
\begin{equation}
    t_s = \frac{\rho\epsilon(1-\epsilon)}{K}.
\end{equation}
Notice that Eq. \ref{eq:DeltaVel} is slightly different from the form presented in \cite{Laibe2014a} due to an error in the original derivation.\\
Although Eq. \ref{eq:DeltaVel} is the most compact form for the evolution of $\vec{\Delta v}$, here we present two alternative formulations. The first formulation is the one derived in \cite{Lebreuilly2019}
\begin{equation}
    \label{eq:DeltaVel_Lebreuilly}
    \begin{split}
        \frac{d\vec{\Delta v}}{dt} & =  \frac{\nabla P}{\rho(1-\epsilon)}-\frac{\vec{\Delta v}}{t_s} - (\vec{\Delta v}\cdot\nabla)\vec{v} + \frac{1}{2}\nabla[(2\epsilon-1)\vec{\Delta v}\cdot \vec{\Delta v}] \\
                                   & + (1-\epsilon)\vec{\Delta v}\times[\nabla\times(1-\epsilon)\vec{\Delta v}] - \epsilon\vec{\Delta v}\times[\nabla\times\epsilon\vec{\Delta v}].
    \end{split}
\end{equation}
The second version will be the one that we will use for the later SPH discretization
\begin{equation}
    \label{eq:DeltaVel_SPHfriendly}
    \begin{split}
         & \frac{d\vec{\Delta v}}{dt} =  \frac{\nabla P}{\rho(1-\epsilon)}-\frac{\vec{\Delta v}}{t_s} - (\vec{\Delta v}\cdot\nabla)\vec{v} + \frac{1}{2}\nabla[(2\epsilon-1)\vec{\Delta v}\cdot \vec{\Delta v}] \\
         & + \vec{\Delta v}(\vec{\Delta v}\cdot \nabla)\epsilon + (2\epsilon-1)(\vec{\Delta v}\cdot \nabla)\vec{\Delta v} - |\vec{\Delta v}|^2\nabla\epsilon - \frac{1}{2}(2\epsilon-1)\nabla|\vec{\Delta v}|^2.
    \end{split}
\end{equation}
While all these three formulations seem very different, they are actually equivalent to one another, and each may be useful under particular circumstances.\\
We note here that Eq. \ref{eq:DeltaVel_Lebreuilly} and Eq. \ref{eq:DeltaVel_SPHfriendly} share a common property: under scalar product with $\vec{\Delta v}$ itself, the second row of both equations cancels out. It is easier to see in Eq. \ref{eq:DeltaVel_Lebreuilly}, where all terms in the second row are the result of cross-products with $\vec{\Delta v}$, and thus perpendicular to it, but the same holds for Eq. \ref{eq:DeltaVel_SPHfriendly} as well. This property will be crucial in deriving a suitable SPH discretization of Eq. \ref{eq:DeltaVel_SPHfriendly}.

\subsection{SPH implementation of the One-Fluid model}
In order to computationally solve the One-Fluid model presented above, we need to find a suitable discretization that conserves mass, momentum, and energy. We will closely follow the derivation of \cite{Laibe2014b} (hereafter LP14b) for the well-known Smoothed Particle Hydrodynamics (SPH) framework, but corrected for the new terms in Equation \ref{eq:DeltaVel_SPHfriendly}. We will also add time-dependent versions of the artificial viscosity and conductivity, a new dust diffusion term, and the explicit form of the equations for a parametrization of the dust fraction.

\subsubsection{Density estimate}
In SPH, the fluid is modeled as a collection of particles, each representing a portion of its mass. Any hydrodynamical quantity at a point in space $\vec{r}$ is then represented by a weighted average of the contribution of all neighboring particles, weighted by a kernel $\mathcal{W}(\vec{r}_a,h_a)$, depending on the distance $\vec{r}_a$ between the particle $a$ and the point $\vec{r}$ and the smoothing length $h_a$ of the particle. The kernel should satisfy several properties: compact support, continuity, radial symmetry, and the following normalization constraint
\begin{equation}
    \int_V \mathcal{W}(|\vec{r}'-\vec{r}_a|,h_a) dV' = 1.
\end{equation}
Some common choices for SPH kernels include the cubic \citep{Monaghan1985} and quintic spline \citep{Morris1996} and the Wedland C2/C4 and C6 kernels \citep{Wendland1995, Dehnen2012}.\\
The density of a given particle is then determined by a weighted sum over the particle's neighbors
\begin{equation}
    \label{eq:SPH_density}
    \rho(\vec{r}_a) = \sum_{b\in NGB}m_b \mathcal{W}(|\vec{r}_a - \vec{r}_b|, h_a),
\end{equation}
where $m_b$ is the mass of the $b$ particle. Eq. \ref{eq:SPH_density} can be shown to be a solution to the continuity equation Eq. \ref{eq:continuity}. The variable smoothing length $h_a$ of each particle is computed to keep the mass inside the kernel volume of each particle fixed. In three dimensions, $h_a$ is computed to fulfill the equation
\begin{equation}
    \frac{4}{3}\pi h_a^3 \rho_a = N_{\text{ngb}} m_a,
\end{equation}
where $N_{\text{ngb}}$ is the chosen number of neighbors.
\subsubsection{Differential operators}
To obtain a suitable discretization of the One-Fluid model equations we will employ the following SPH differential operators \citep{Price2012}
\begin{equation}
    \label{SPH_div}
    \nabla\cdot\vec{A} \underset{\text{SPH}}{=} \rho_a \sum_b m_b \left[\frac{\vec{A}_a\cdot\nabla\mathcal{W}_{ab}(h_a)}{\Omega_a\rho_a^2}+\frac{\vec{A}_b\cdot\nabla\mathcal{W}_{ab}(h_b)}{\Omega_b\rho_b^2}\right],
\end{equation}
\begin{equation}
    \label{SPH_A_dot_nabla_f}
    (\vec{A}\cdot \nabla)f \underset{\text{SPH}}{=} \frac{1}{\rho_a\Omega_a}\sum_b m_b (f_b - f_a)\vec{A}_a\cdot \nabla\mathcal{W}_{ab}(h_a),
\end{equation}
\begin{equation}
    \label{SPH_A_dot_nabla_B}
    (\vec{A}\cdot \nabla)\vec{B} \underset{\text{SPH}}{=} \frac{1}{\rho_a\Omega_a}\sum_b m_b (\vec{B}_b - \vec{B}_a)\vec{A}_a\cdot \nabla\mathcal{W}_{ab}(h_a),
\end{equation}
where $\mathcal{W}_{ab}(h_a) = \mathcal{W}(|\vec{r}_a-\vec{r}_b|,h_a)$, $\Omega_a = \left(1 + \frac{h_a}{\rho_ad}\frac{\partial \rho_a}{\partial h_a}\right)$, with $d$ the number of dimensions, and $\vec{A}$, $\vec{B}$ and $f$ are arbitrary vector and scalar fields.\\
We will apply these operators and, step by step, ensure that mass, total momentum, and total energy are conserved.

\subsubsection{Mass conservation}
The first requirement that the SPH discretization should meet is the conservation of mass. The total mass of the mixture is conserved by construction by the SPH framework, and since there is no dust-mass evolution mechanism taken into account in the One-Fluid model (e.g., sputtering, erosion, or astration), the individual dust and gas masses should also be conserved. Dust mass conservation requires that
\begin{equation}
    \frac{dM_d}{dt} \underset{\text{SPH}}{=} \sum_a m_a \frac{d\epsilon_a}{dt} = 0.
\end{equation}
By applying the divergence operator Eq. \ref{SPH_div} to Eq. \ref{eq:dust_frac} we obtain the following SPH discretization
\begin{equation}
    \begin{split}
        \frac{d\epsilon_a}{dt} = -\sum_b m_b \Biggl[ \frac{\epsilon_a(1-\epsilon_a)}{\Omega_a\rho_a} & \vec{\Delta v}_a\cdot\nabla_a\mathcal{W}_{ab}(h_a)                                                          \\
                                                                                                     & + \frac{\epsilon_b(1-\epsilon_b)}{\Omega_b\rho_b}\vec{\Delta v}_b\cdot\nabla_a\mathcal{W}_{ab}(h_b) \Biggr].
    \end{split}
    \label{eq:dEpsilondt_SPH}
\end{equation}
Using the anti-symmetry of the kernel gradient $\nabla\mathcal{W}_{ab}(h_a) = -\nabla\mathcal{W}_{ab}(h_b)$, it is straightforward to show that the total mass of dust is conserved.

\subsubsection{Momentum conservation}
While dust and gas can exchange momentum, the total momentum of the mixture should be conserved. A suitable SPH discretization of Eq. \ref{eq:momentum} is constrained by
\begin{equation}
    \sum_a m_a \frac{d\vec{v}_a}{dt} = 0.
    \label{eq:momentum_conservation}
\end{equation}
Compared to the gas-only momentum conservation equation, the One-Fluid model version, Eq. \ref{eq:momentum}, has a first term that represents the usual pressure gradient acceleration and a new term specific to the One-Fluid model. The pressure gradient term can be solved as usual in SPH
\begin{align}
    \biggl(\frac{d\vec{v}_a}{dt}\biggr)_{\nabla P} = -\sum_b m_b  \biggl[\frac{P_a}{\Omega_a \rho_a^2}\nabla\mathcal{W}_{ab}(h_a) + \frac{P_b}{\Omega_b \rho_b^2}\nabla\mathcal{W}_{ab}(h_b) \biggr].
    \label{eq:pressure_gradient_force}
\end{align}
The Eq. \ref{eq:DeltaVel_SPHfriendly} also presents a similar term, which we discretize in the same manner
\begin{equation}
    \begin{split}
        \biggl(\frac{d\vec{\Delta v}_a}{dt}\biggr)_{\nabla P} & = -\frac{1}{1-\epsilon_a} \biggl(\frac{d\vec{v}_a}{dt}\biggr)_{\nabla P} =                                                                                            \\
                                                              & =\frac{1}{1-\epsilon_a}\sum_b m_b  \biggl[\frac{P_a}{\Omega_a \rho_a^2}\nabla\mathcal{W}_{ab}(h_a) + \frac{P_b}{\Omega_b \rho_b^2}\nabla\mathcal{W}_{ab}(h_b) \biggr].
    \end{split}
    \label{eq:pressure_gradient_force_DeltaVel}
\end{equation}
Back to the momentum conservation, the second term of Eq. \ref{eq:momentum} can be discretized by applying the differential operator Eq. \ref{SPH_div}
\begin{equation}
    \begin{split}
        \biggl(\frac{d\vec{v}_a}{dt}\biggr)_{OFM} = - \sum_b m_b \biggl[ & \frac{\epsilon_a(1-\epsilon_a)\vec{\Delta v}_a}{\Omega_a \rho_a}\vec{\Delta v}_a \cdot \nabla\mathcal{W}_{ab}(h_a) +       \\
                                                                         & +\frac{\epsilon_b(1-\epsilon_b)\vec{\Delta v}_b}{\Omega_b \rho_b}\vec{\Delta v}_b \cdot \nabla\mathcal{W}_{ab}(h_b)\biggr].
        \label{eq:momentum_OFM}
    \end{split}
\end{equation}
Both Eq. \ref{eq:pressure_gradient_force}  and Eq. \ref{eq:momentum_OFM} conserve momentum as required by Eq. \ref{eq:momentum_conservation} and can thus be used as SPH discretizations of the corresponding continuum equations.

\subsubsection{Energy conservation}
The mixture of dust and gas should also conserve its total energy, and from this requirement, we will derive the SPH discretizations for both the internal energy $u$ and the remaining terms in the $\vec{\Delta v}$ evolution.\\
The total energy of the mixture is the sum of the total energy of the gas phase (kinetic and internal energy) and the total energy of the dust phase (which only has a kinetic component, as it is modeled as a pressureless fluid)
\begin{equation}
    \begin{split}
        E & = \int_V \biggl(\frac{1}{2}m_g\vec{v}_g^2 + m_g u + \frac{1}{2}m_d\vec{v}_d^2\biggr) dV =                                                             \\
          & = \int_V m \biggl(\frac{1}{2}\vec{v}^2 + (1-\epsilon)u + \frac{1}{2}\epsilon(1-\epsilon)\vec{\Delta v}^2\biggr) dV \underset{\text{SPH}}{=}           \\
          & \underset{\text{SPH}}{=} \sum_a m_a \biggl(\frac{1}{2}\vec{v}_a^2 + (1-\epsilon_a)u_a  + \frac{1}{2}\epsilon_a(1-\epsilon_a)\vec{\Delta v}_a^2\biggr),
    \end{split}
    \label{eq:total_energy}
\end{equation}
where in the second row we introduced the One-Fluid model variables and in the final row we discretized the integral as a sum over all particles.\\
By computing the time derivative of Eq. \ref{eq:total_energy} we obtain the constraint that the SPH discretizations for Eq. \ref{eq:DeltaVel_SPHfriendly} and \ref{eq:energy} should satisfy
\begin{equation}
    \begin{split}
        \frac{dE}{dt} = \sum_a m_a & \biggl[\vec{v}_a\cdot\frac{d\vec{v}_a}{dt} + \epsilon_a(1-\epsilon_a)\vec{\Delta v}_a\cdot \frac{d\vec{\Delta v}_a}{dt} +         \\
                                   & \biggl((1-2\epsilon_a)\frac{\vec{\Delta v}_a^2}{2} - u_a\biggr) \frac{d\epsilon_a}{dt} + (1-\epsilon_a)\frac{du_a}{dt}\biggr] = 0.
    \end{split}
    \label{eq:energy_conservation}
\end{equation}
The first term we want to discretize is the $PdV$ term in Eq. \ref{eq:energy}. By combining all terms in which pressure is present, we obtain the following constraint from energy conservation
\begin{equation}
    \begin{split}
        \sum_a m_a & (1-\epsilon_a)\biggl(\frac{du_a}{dt}\biggr)_{PdV} =                                                                                                                                               \\
                   & =- \sum_a m_a \biggl[\vec{v}_a\cdot \biggl(\frac{d\vec{v}_a}{dt}\biggr)_{\nabla P} + \epsilon_a(1-\epsilon_a)\vec{\Delta v}_a\cdot \biggl(\frac{d\vec{\Delta v}_a}{dt}\biggr)_{\nabla P}\biggr] = \\
                   & = -\sum_a m_a \biggl[(\vec{v}_a - \epsilon_a \vec{\Delta v}_a) \cdot \biggl(\frac{d\vec{v}_a}{dt}\biggr)_{\nabla P} \biggr]                                                                       \\
                   & =\sum_a m_a \sum_b m_b \frac{P_a} {\Omega_a\rho_a^2}(\vec{v}_a-\epsilon_a\vec{\Delta v}_a) \cdot \nabla\mathcal{W}_{ab}(h_a) +                                                                    \\
                   & \qquad+ \sum_a m_a \sum_b m_b \frac{P_b}{\Omega_b\rho_b^2}(\vec{v}_a-\epsilon_a\vec{\Delta v}_a) \cdot \nabla\mathcal{W}_{ab}(h_b),
    \end{split}
    \label{energy_conservation_equivalence}
\end{equation}
where in the last step we have substituted $\biggl(\frac{d\vec{v}_a}{dt}\biggr)_{\nabla P}$ from Eq. \ref{eq:pressure_gradient_force}. Swapping the summation indices of the second term and using the anti-symmetry of the kernel, we obtain
\begin{equation}
    \biggl(\frac{du_a}{dt}\biggr)_{PdV} = \frac{P_a}{\Omega_a \rho_a^2 (1-\epsilon_a)}\sum_b m_b[(\vec{v}_a-\epsilon_a\vec{\Delta v}_a - \vec{v}_b + \epsilon_b \vec{\Delta v}_b) \cdot \mathcal\nabla{W}_{ab}(h_a)].
\end{equation}
The second term in Eq. \ref{eq:energy} is also constrained from energy conservation. By comparing the terms in Eq. \ref{eq:energy_conservation} involving the internal energy
\begin{equation}
    \sum_a m_a (1-\epsilon_a) \biggl(\frac{du_a}{dt}\biggr)_{(\vec{\Delta v}\cdot \nabla)u} = \sum_a m_a u_a \frac{d\epsilon_a}{dt}.
\end{equation}
By substituting Eq. \ref{eq:dEpsilondt_SPH} and swapping indices in the double summation, we obtain
\begin{equation}
    \biggl(\frac{du_a}{dt}\biggr)_{(\vec{\Delta v}\cdot \nabla)u} = \frac{\epsilon_a}{\Omega_a \rho_a}\sum_b m_b (u_b - u_a) \vec{\Delta v}_a\cdot \nabla \mathcal{W}_{ab}(h_a).
\end{equation}
We will now deal with Eq. \ref{eq:DeltaVel_SPHfriendly}, describing the evolution of $\vec{\Delta v}$. The first term, similar to the pressure gradient force term, was already dealt with and its SPH discretization is described in Eq. \ref{eq:pressure_gradient_force_DeltaVel}.
The second term in Equation \ref{eq:DeltaVel_SPHfriendly} relating to the drag will be integrated implicitly and addressed later, together with the last term in Equation \ref{eq:energy}. The SPH discretization for the third term in Eq. \ref{eq:DeltaVel_SPHfriendly} can be derived from energy conservation, by equating the first two terms in Eq. \ref{eq:energy_conservation}:
\begin{equation}
    \sum_a m_a \epsilon_a(1-\epsilon_a)\vec{\Delta v}_a\cdot \biggl(\frac{d\vec{\Delta v}_a}{dt}\biggr)_{-(\vec{\Delta v}\cdot \nabla)\vec{v}} = - \sum_a m_a \vec{v}_a\cdot \biggl(\frac{d\vec{v}_a}{dt}\biggr)_{OFM}.
\end{equation}
Also in this case, we simply substitute Eq. \ref{eq:momentum_OFM} and swap summation indices to obtain
\begin{equation}
    \biggl(\frac{d\vec{\Delta v}_a}{dt}\biggr)_{-(\vec{\Delta v}\cdot \nabla)\vec{v}} = -\frac{1}{\Omega_a \rho_a}\sum_b m_b [(\vec{v}_b - \vec{v}_a) \vec{\Delta v}_a\cdot \nabla \mathcal{W}_{ab}(h_a)].
\end{equation}
The fourth term in Eq. \ref{eq:DeltaVel_SPHfriendly} is also constrained by energy conservation, but comparing now the second and third terms in Eq. \ref{eq:energy_conservation}
\begin{equation}
    \sum_a m_a \epsilon_a(1-\epsilon_a)\vec{\Delta v}_a\cdot \biggl(\frac{d\vec{\Delta v}_a}{dt}\biggr)_{4} = - \sum_a m_a (1-2\epsilon_a)\frac{\vec{\Delta v}_a^2}{2} \frac{d\epsilon_a}{dt}.
\end{equation}
By substituting Eq. \ref{eq:dEpsilondt_SPH} and swapping summation indices we obtain
\begin{equation}
    \begin{split}
        \biggl(\frac{d\vec{\Delta v}_a}{dt}\biggr)_{4} = \frac{1}{2\Omega_a \rho_a}\sum_b m_b [ & (1-2\epsilon_a)\vec{\Delta v}_a^2 - \\&-(1-2\epsilon_b)\vec{\Delta v}_b^2] \nabla \mathcal{W}_{ab}(h_a).
    \end{split}
\end{equation}
It is crucial to note here that the remaining terms in Eq. \ref{eq:DeltaVel_SPHfriendly} cannot be constrained by energy conservation, as they cancel out when multiplied scalarly by $\vec{\Delta v}$, as noted when they were first introduced (see Section \ref{sec:2.1}). Hence, their SPH discretization will be obtained directly by use of SPH differential operators (Eq. \ref{SPH_div} - \ref{SPH_A_dot_nabla_B}), and we will only later check that energy is conserved. We propose the following SPH discretizations
\begin{equation}
    \vec{\Delta v}(\vec{\Delta v}\cdot \nabla)\epsilon \underset{\text{SPH}}{=} \frac{\vec{\Delta v}_a}{\Omega_a \rho_a}\sum_b m_b (\epsilon_b - \epsilon_a)\vec{\Delta v}_a \cdot \nabla \mathcal{W}_{ab}(h_a),
\end{equation}
\begin{equation}
    (2\epsilon-1)(\vec{\Delta v}\cdot\nabla)\vec{\Delta v} \underset{\text{SPH}}{=} \frac{2\epsilon_a-1}{\Omega_a \rho_a}\sum_b m_b (\vec{\Delta v}_b - \vec{\Delta v}_a) \vec{\Delta v}_a\cdot \nabla \mathcal{W}_{ab}(h_a),
\end{equation}
\begin{equation}
    -|\vec{\Delta v}|^2\nabla \epsilon \underset{\text{SPH}}{=} - \frac{|\vec{\Delta v}_a|^2}{\Omega_a \rho_a} \sum_b m_b (\epsilon_b - \epsilon_a) \nabla \mathcal{W}_{ab}(h_a),
\end{equation}
\begin{equation}
    -\frac{1}{2}(2\epsilon-1)\nabla |\vec{\Delta v}|^2 \underset{\text{SPH}}{=} -\frac{2\epsilon_a-1}{\Omega_a\rho_a} \sum_b m_b (\vec{\Delta v}_b - \vec{\Delta v}_a)\cdot \vec{\Delta v}_a\nabla\mathcal{W}_{ab}(h_a).
\end{equation}
It can be shown that these SPH discretizations conserve energy if inserted into Eq. \ref{eq:energy_conservation}.\\
One quantity that this set of SPH discretizations fails to conserve is the dust momentum in the absence of drag forces. However, deviations from conservation are only of order $\mathcal{O}(\vec{\Delta v}^2)$, and thus negligible for strong drag. In the weak drag case, we also observed little deviation from conservation in the benchmark tests we show in Section 3. In general, the conservation of dust momentum for zero drag is a property that has still not been enforced by previous SPH discretizations of the One-Fluid model, and could be an area of future improvement.

\subsection{Timestepping}
\texttt{OpenGadget3} employs, for integrating in time all hydrodynamical variables, a leapfrog scheme in a Kick-Drift-Kick (KDK) form. This scheme ensures second-order accuracy for all variables and is implemented as in \cite{Springel2005}.\\
We indicate with $\vec{W} = (\epsilon, \vec{v}, \vec{\Delta v}, u)$ the vector of primitive quantities of the One-Fluid model, except for density, which is computed by the kernel density estimate in Eq. \ref{eq:SPH_density}. The $\vec{W}$ vector and the positions of the particles, $\vec{x}$, are evolved using predicted quantities to compute the increments of the second kick
\begin{equation}
    \begin{split}
         & \vec{W}_{n+1/2} = \vec{W}_n + \frac{1}{2}\Delta t~ \biggl(\frac{d\vec{W}}{dt}\biggr)_n,           \\
         & \vec{W}_{n+1}^* = \vec{W}_n + \Delta t~ \biggl(\frac{d\vec{W}}{dt}\biggr)_n,                      \\
         & \vec{x}_{n+1} = \vec{x}_n + \Delta t~\vec{v}_{n+1/2},                                            \\
         & \vec{W}_{n+1} = \vec{W}_{n+1/2} + \frac{1}{2}\Delta t~ \biggl(\frac{d\vec{W}}{dt}\biggr)_{n+1}^*.
    \end{split}
\end{equation}

\subsection{Implicit integration of the drag term}
The last two terms in Eq. \ref{eq:DeltaVel_SPHfriendly} and Eq. \ref{eq:energy} can be numerically integrated both explicitly and implicitly\footnote{Technically, the integrator presented here should be referred to as "analytic", rather than "implicit". However, in practice, all integrators of drag that are not explicit are usually referred to as "implicit" in the literature.}. As shown in \citetalias{Laibe2014b}, explicit integration is subject to the timestep requirement $\Delta t < t_s$, which becomes prohibitive for strong drag. Implicit integration, instead, does not require such a timestep requirement and can be implemented via operator splitting.\\
We show here the procedure for the first kick in the KDK scheme, but the same is applied also to compute the predicted quantities during the drift and for the second kick. By separating the drag term from all the other contributions, we can rewrite Eq. \ref{eq:DeltaVel_SPHfriendly} and \ref{eq:energy} as
\begin{equation}
    \begin{cases}
        \displaystyle\frac{d\vec{\Delta v}}{dt} = \biggl(\displaystyle\frac{d\vec{\Delta v}}{dt}\biggr)_0 - \displaystyle\frac{\vec{\Delta v}}{t_s}, \\[10pt]
        \displaystyle\frac{du}{dt} = \biggl(\displaystyle\frac{du}{dt}\biggr)_0 + \epsilon\displaystyle\frac{\vec{\Delta v}^2}{t_s}.
    \end{cases}
    \label{eq:system}
\end{equation}
To use operator splitting, we first solve the corresponding homogeneous system by ignoring the drag terms and using the SPH discretization derived above
\begin{equation}
    \begin{cases}
        \overline{\vec{\Delta v}}_{n+1/2} = \vec{\Delta v}_{n} + \displaystyle\frac{1}{2}\Delta t~ \biggl(\displaystyle\frac{d\vec{\Delta v}}{dt}\biggr)_{n,0}, \\[10pt]
        \overline{u}_{n+1/2} = u_n + \displaystyle\frac{1}{2}\Delta t ~\biggl(\displaystyle\frac{du}{dt} \biggr)_{n,0}.
    \end{cases}
\end{equation}
The Ordinary Differential Equation corresponding to the system in Eq. \ref{eq:system} can be solved analytically as
\[
    \begin{array}{c}
        \begin{cases}
            \displaystyle\frac{d\vec{\Delta v}}{dt} = -\displaystyle\frac{\vec{\Delta v}}{t_s}, \\[10pt]
            \displaystyle\frac{du}{dt}  = \epsilon \displaystyle\frac{\vec{\Delta v}^2}{t_s},
        \end{cases}
        \quad \xrightarrow{} \quad
        \begin{cases}
            \vec{\Delta v}(t) = \vec{\Delta v}_0 e^{-\frac{t}{t_s}}, \\[10pt]
            u (t) = u_0 - \displaystyle\frac{1}{2}\epsilon\vec{\Delta v}^2_0\left[e^{-\frac{2t}{t_s}}-1\right].
        \end{cases}
    \end{array}
\]
By discretizing this solution to half a timestep as in the first kick, and by using the solutions to the homogeneous system as initial conditions, we obtain the final result
\begin{equation}
    \begin{cases}
        \vec{\Delta v}_{n+1/2} = \left[\vec{\Delta v}_{n} + \displaystyle\frac{1}{2}\Delta t
        \biggl(\displaystyle\frac{d\vec{\Delta v}}{dt}\biggr)_{n,0}\right]e^{-\frac{\Delta t}{2t_s}},                                                                                                                \\[10pt]
        u_{n+1/2} = u_n + \displaystyle\frac{1}{2}\Delta t \biggl(\displaystyle\frac{du}{dt} \biggr)_{n,0} - \frac{1}{2}\epsilon_{n+1/2}\overline{\vec{\Delta v}}^2_{n+1/2}\left[e^{-\frac{\Delta t}{t_s}}-1\right]. \\
    \end{cases}
\end{equation}

\subsection{Time-dependent Artificial Viscosity}
In order to dampen post-shock oscillations, prevent particle interpenetration, and treat contact discontinuities, SPH frameworks often employ artificial viscosity mechanisms \citep{Price2008}. Our artificial viscosity implementation follows the steps of \citetalias{Laibe2014b}, with some modifications and improvements that make the implementation time-dependent. We show both the conservative (C) and the non-conservative (NC) approaches to artificial viscosity, as shown in \citetalias{Laibe2014b}, an artificial dissipation term for $\vec{\Delta v}$, and a simple unifying scheme to combine both conservative and non-conservative viscosities.
\subsubsection{Conservative Artificial Viscosity}
A simple momentum-conserving formulation for artificial viscosity can be written as
\begin{equation}
    \biggl(\frac{d\vec{v}}{dt}\biggr)_{AV}^C = \sum_b m_b (1-\overline{\epsilon}_{ab}) \nu_{ab} \overline{\nabla \mathcal{W}}_{ab},
    \label{dvdtCAV}
\end{equation}
where the overline stands for the mean value between particle $a$ and particle $b$ of a given quantity. For instance, $\overline{\epsilon}_{ab} = \frac{1}{2}(\epsilon_a + \epsilon_b)$.\\
A possible choice for the viscosity $\nu_{ab}$ is the one employed by \citetalias{Laibe2014b}
\begin{equation}
    \nu_{ab} = \frac{v^{sig}_{ab}}{\overline{\rho}_{ab}}\vec{v}_{ab}^g\cdot \hat{\vec{r}}_{ab},
\end{equation}
where $\vec{v}_{ab}^g = \vec{v}_{a}^g -\vec{v}_b^g$ is the difference between the gas velocities of the particles, $\hat{\vec{r}}_{ab}$ is the unit vector pointing from particle $a$ to particle $b$ and $v^{sig}_{ab}$ is the signal velocity.\\
The signal velocity is used to switch on and off the artificial viscosity depending on whether the gas in the particles is approaching or distancing
\begin{equation}
    v^{sig}_{ab} = \begin{cases}
        \frac{1}{2}(c_{s,a} + c_{s,b} - \beta \vec{v}_{ab}^g \cdot  \hat{\vec{r}}_{ab}) & \vec{v}_{ab}^g \cdot  \hat{\vec{r}}_{ab} \leq 0, \\
        0                                                                               & \vec{v}_{ab}^g \cdot  \hat{\vec{r}}_{ab} > 0,
    \end{cases}
    \label{eq:signal_velocity}
\end{equation}
where $c_s$ is the sound speed of the particle and $\beta = 3$ following \cite{Beck2015}.\\
In our implementation, we adopt two additional terms in the formulation of viscosity
\begin{equation}
    \nu_{ab} = \frac{v^{sig}_{ab}}{\overline{\rho}_{ab}} \alpha_{ab} \vec{v}_{ab}^g\cdot \hat{\vec{r}}_{ab},
    \label{eq:viscosity}
\end{equation}
where $\alpha_{ab}$ is the symmetrized viscosity coefficient, and is used to reduce artificial viscosity when it is not needed, i.e., away from shocks.
This framework was already implemented \citep{Dolag2005, Beck2015} and extensively tested in \texttt{OpenGadget3} by \cite{Marin-Gilabert2022}, we only added the precaution of using the gas velocity instead of the mixture velocity when computing $\alpha_{ab}$.\\
In order to make this formulation conservative, we need to add matching terms in the equations of $\vec{\Delta v}$ and $u$. In particular, for $\vec{\Delta v}$ we simply have
\begin{equation}
    \biggl(\frac{d\vec{\Delta v}_a}{dt}\biggr)_{AV}^C = \frac{1}{1-\epsilon_a}\sum_b m_b (1-\overline{\epsilon}_{ab}) \nu_{ab} \overline{\nabla \mathcal{W}}_{ab},
    \label{ddvdtCAV}
\end{equation}
while the term in the internal energy equation is constrained from the equivalence already shown in Eq. \ref{energy_conservation_equivalence}
\begin{equation}
    \sum_a m_a (1-\epsilon_a)\biggl(\frac{du_a}{dt}\biggr)_{AV}^C = -\sum_a m_a \biggl[\vec{v}_a^g \cdot \biggl(\frac{d\vec{v}_a}{dt}\biggr)_{AV}^C \biggr],
\end{equation}
which gives, after inserting Eq. \ref{dvdtCAV}, expanding $\overline{\nabla\mathcal{W}_{ab}}$ and swapping summation indices
\begin{equation}
    \biggl(\frac{du_a}{dt}\biggr)_{AV}^C = -\frac{1}{2(1-\epsilon_a)}\sum_b m_b (1-\overline{\epsilon}_{ab}) \nu_{ab} \vec{v}_{ab}^g \cdot \hat{\vec{r}}_{ab} \overline{F}_{ab},
    \label{dudtCAV}
\end{equation}
where the kernel gradient has been expanded into direction and magnitude: $\nabla\mathcal{W}_{ab}(h_a) = \hat{\vec{r}}_{ab} F_{ab}(h_a)$.\\
As described by \citetalias{Laibe2014b}, only using Eq. \ref{dvdtCAV}, \ref{ddvdtCAV} and \ref{dudtCAV} leaves some oscillations in $\vec{\Delta v}$ in post-shock regions, which can be treated by using an additional dissipative term in the $\vec{\Delta v}$ equation. In particular, a momentum and energy-conserving formulation is
\begin{equation}
    \biggl(\frac{d\vec{\Delta v}_a}{dt}\biggr)_{AD} = \frac{1}{\epsilon_a(1-\epsilon_a)}\sum_b\frac{v_{sig}^{\vec{\Delta v}}}{\overline{\rho}_{ab}} \overline{\epsilon}_{ab}(1-\overline{\epsilon}_{ab})\vec{\Delta v}_{ab}\cdot\hat{\vec{r}}_{ab} \overline{\nabla \mathcal{W}_{ab}}.
\end{equation}
In order to conserve energy, a matching term in the internal energy equation must be added
\begin{equation}
    \biggl(\frac{du_a}{dt}\biggr)_{AD} = -\frac{1}{2(1-\epsilon_a)}\sum_b m_b \frac{v_{sig}^{\Delta v}}{\overline{\rho}_{ab}}\overline{\epsilon}_{ab}(1-\overline{\epsilon}_{ab})(\vec{\Delta v}_{ab}\cdot\hat{\vec{r}}_{ab})^2 \overline{F}_{ab}.
\end{equation}
The signal velocity used here is different from the one used in the artificial viscosity. In particular, \citetalias{Laibe2014b} propose the following definition
\begin{equation}
    v^{sig, \Delta v}_{ab} = \begin{cases}
        - \alpha_{\Delta v} \vec{\Delta v}_{ab}\cdot\hat{\vec{r}}_{ab} & \vec{\Delta v}_{ab} \cdot  \hat{\vec{r}}_{ab} \leq 0, \\
        0                                                              & \vec{\Delta v}_{ab} \cdot  \hat{\vec{r}}_{ab} > 0,
    \end{cases}
\end{equation}
with $\alpha_{\Delta v} = 1$.
This concludes the conservative formulation for artificial viscosity. As already found by \citetalias{Laibe2014b}, this formulation works well for low $\vec{\Delta v}$, but fails in shocks with large $\vec{\Delta v}$, as can be seen in the DUSTYSHOCK benchmark test, at low drag. For this reason, \citetalias{Laibe2014b} introduced an alternative, non-conservative, formulation of artificial viscosity that shows a much better performance on the DUSTYSHOCK test.

\subsubsection{Nonconservative Artificial Viscosity}
The nonconservative formulation of artificial viscosity we employed is very similar to the one derived by \citetalias{Laibe2014b}, with the only improvement of making it adaptive and time-dependent by using the viscosity in Eq. \ref{eq:viscosity}
\begin{equation}
    \biggl(\frac{d\vec{v}_a}{dt}\biggr)_{AV}^{NC} = \sum_b m_b  \nu_{ab} \overline{\nabla \mathcal{W}}_{ab},
    \label{dvdtNCAV}
\end{equation}
\begin{equation}
    \biggl(\frac{d\vec{\Delta v}_a}{dt}\biggr)_{AV}^{NC} = \frac{1}{1-\epsilon_a}\sum_b\frac{v^{sig}_{ab}}{\overline{\rho}_{ab}} \vec{\Delta v}_{ab}\cdot\hat{\vec{r}}_{ab} \overline{\nabla \mathcal{W}_{ab}},
    \label{ddvdtNCAV}
\end{equation}
where the signal velocity employed in the second equation is the one in Eq. \ref{eq:signal_velocity}.\\
The viscous heating term is also added as in Eq. \ref{dudtCAV}, slightly modified to adjust to the different term in the $\frac{d\vec{v}}{dt}$ equation
\begin{equation}
    \biggl(\frac{du_a}{dt}\biggr)_{AV}^{NC} = -\frac{1}{2(1-\epsilon_a)}\sum_b m_b \nu_{ab} \vec{v}_{ab}^g \cdot \hat{\vec{r}}_{ab} \overline{F}_{ab}(h_a).
    \label{dudtNCAV}
\end{equation}

\subsubsection{Unified framework}
In order to derive a unified framework that could be used under all circumstances, instead of relying on two options to be left to the user, we use the shock indicator $R_{a}$ to interpolate between the two formulations. The shock indicator is defined as in \cite{Cullen2010}
\begin{equation}
    R_{a} = \frac{1}{\rho_a}\sum_b \text{sign}(\nabla\cdot\vec{v}_b)m_b \mathcal{W}_{ab}.
\end{equation}
For non-shocked regions, $R_{a}\simeq 0$, while for shocked regions $R_{a} \simeq\pm 1$. This indicator is already used to compute the symmetrized viscosity coefficient $\alpha_{ab}$, to determine the strength of the shock.\\
Starting from the shock indicator, we define a new variable, $\sigma_{ab}$, to interpolate between the conservative and the non-conservative formulations of artificial viscosity. $\sigma_{ab}$ is defined as a sigmoid function
\begin{equation}
    \sigma_{ab} = \frac{1}{1 + e^{\alpha(\overline{|R|}_{ab} - \beta)}},
\end{equation}
where appropriate values for $\alpha$ and $\beta$ where found to be $\alpha = 20$ and $\beta = 0.25$. This makes the sigmoid function steep enough to quickly transition from one formulation to the other. In particular, for shocked regions $\sigma_{ab} \simeq 0$, while for non-shocked regions $\sigma_{ab} \simeq 1$.\\
The interpolation between conservative and non-conservative formulations is performed as follows
\begin{equation}
    \biggl(\frac{d\vec{v}}{dt}\biggr)_{AV} = \biggl(\frac{d\vec{v}}{dt}\biggr)_{AV}^C \sigma_{ab} + \biggl(\frac{d\vec{v}}{dt}\biggr)_{AV}^{NC} (1-\sigma_{ab}),
\end{equation}
\begin{equation}
    \biggl(\frac{d\Delta\vec{v}}{dt}\biggr)_{AV} = \biggl[\biggl(\frac{d\Delta\vec{v}}{dt}\biggr)_{AV}^C + \biggl(\frac{d\Delta\vec{v}}{dt}\biggr)_{AD}  \biggr]\sigma_{ab} + \biggl(\frac{d\Delta \vec{v}}{dt}\biggr)_{AV}^{NC} (1-\sigma_{ab}),
\end{equation}
\begin{equation}
    \biggl(\frac{du}{dt}\biggr)_{AV} =\biggl[ \biggl(\frac{du}{dt}\biggr)_{AV}^C + \biggl(\frac{du}{dt}\biggr)_{AD} \biggr] \sigma_{ab} + \biggl(\frac{du}{dt}\biggr)_{AV}^{NC} (1-\sigma_{ab}).
\end{equation}
This is the framework that will be used throughout all the benchmark tests and simulations.

\subsection{Time-dependent Artificial Conductivity}
Artificial conductivity is introduced to treat discontinuities in internal energy \citep{Price2008}. Here we employ its time-dependent formulation introduced in \cite{Beck2015}, which is also used in \texttt{OpenGadget3} \citep[see][]{Marin-Gilabert2022,Groth2023}, adapted for the One-Fluid model.,
\begin{equation}
    \biggl(\frac{du_a}{dt}\biggr)_{AC} = \frac{1}{1-\epsilon_a}\sum_b \frac{m_b}{\overline{\rho}_{ab}} (u_b - u_a) \alpha_{ab}^C v_{ab}^{sig,c}\overline{F}_{ab},
\end{equation}
where the signal velocity for the artificial conductivity is defined as
\begin{equation}
    v_{ab}^{sig,c} = \sqrt{\frac{|P_a - P_b|}{\overline{\rho(1-\epsilon)}_{ab}}},
\end{equation}
and the conductivity coefficient $\alpha_{a}^C$ is made time-dependent by computing it from the internal energy and its gradient:
\begin{equation}
    \alpha_{a}^C = \frac{h_a}{2}\frac{|\nabla u_a|}{|u_a|}.
\end{equation}

\subsection{Dust fraction parametrization}
In order to ensure that the dust fraction, $\epsilon$, remains bounded between $0$ and $1$, a suitable parametrization is needed. Here we follow \citep{Ballabio2018} and implement the following parametrization
\begin{equation}
    \epsilon = \frac{s^2}{1+s^2} \rightarrow s = \sqrt{\frac{\epsilon}{1 - \epsilon}}.
    \label{eq:dust_frac_param}
\end{equation}
The corresponding continuum equation for the variable $s$ can be obtained by substituting the parametrization in Eq. \ref{eq:dust_frac} and by using that
\begin{equation}
    \frac{ds}{dt} = \frac{(1+s^2)^2}{2s}\frac{d\epsilon}{dt},
\end{equation}
which results in
\begin{equation}
    \frac{ds}{dt} = -\frac{(1+s^2)^2}{2\rho s}\nabla \cdot \biggl[\frac{s^2}{(1+s^2)^2}~\rho\vec{\Delta v}\biggr].
\end{equation}
Since the $s$ at the denominator could cause numerical issues if a given SPH particle has very little or no dust, we split the $s^2 = s \cdot s$ inside the divergence operator to obtain the following
\begin{equation}
    \frac{ds}{dt} = - \frac{(1+s^2)^2}{2\rho}\nabla\cdot \biggl[\frac{s}{(1+s^2)^2}~\rho\vec{\Delta v}\biggr] - \frac{1}{2}(\vec{\Delta v} \cdot \nabla) s.
\end{equation}
This last formulation is thus discretized using the SPH operators Eq. \ref{SPH_div}-\ref{SPH_A_dot_nabla_f} and rearranged to obtain
\begin{equation}
    \begin{split}
        \frac{ds_a}{dt} = -\frac{1}{2}\sum_b m_b s_b & \biggl[\frac{1}{\Omega_a \rho_a}\vec{\Delta v}_a\cdot \nabla \mathcal{W}_{ab}(h_a)                        \\
                                                     & + \frac{(1+s_a^2)^2}{\Omega_b \rho_b(1+s_b^2)^2}\vec{\Delta v}_b\cdot \nabla \mathcal{W}_{ab}(h_b)\biggr].
    \end{split}
\end{equation}
The other equations and corresponding SPH discretizations do not need to be recomputed; we simply replace $\epsilon$ when needed with the parametrization in Eq. \ref{eq:dust_frac_param}, in terms of the new variable $s$.

\subsection{Artificial Diffusion of dust}
\begin{figure}
    \centering
    \includegraphics[width=0.8\linewidth]{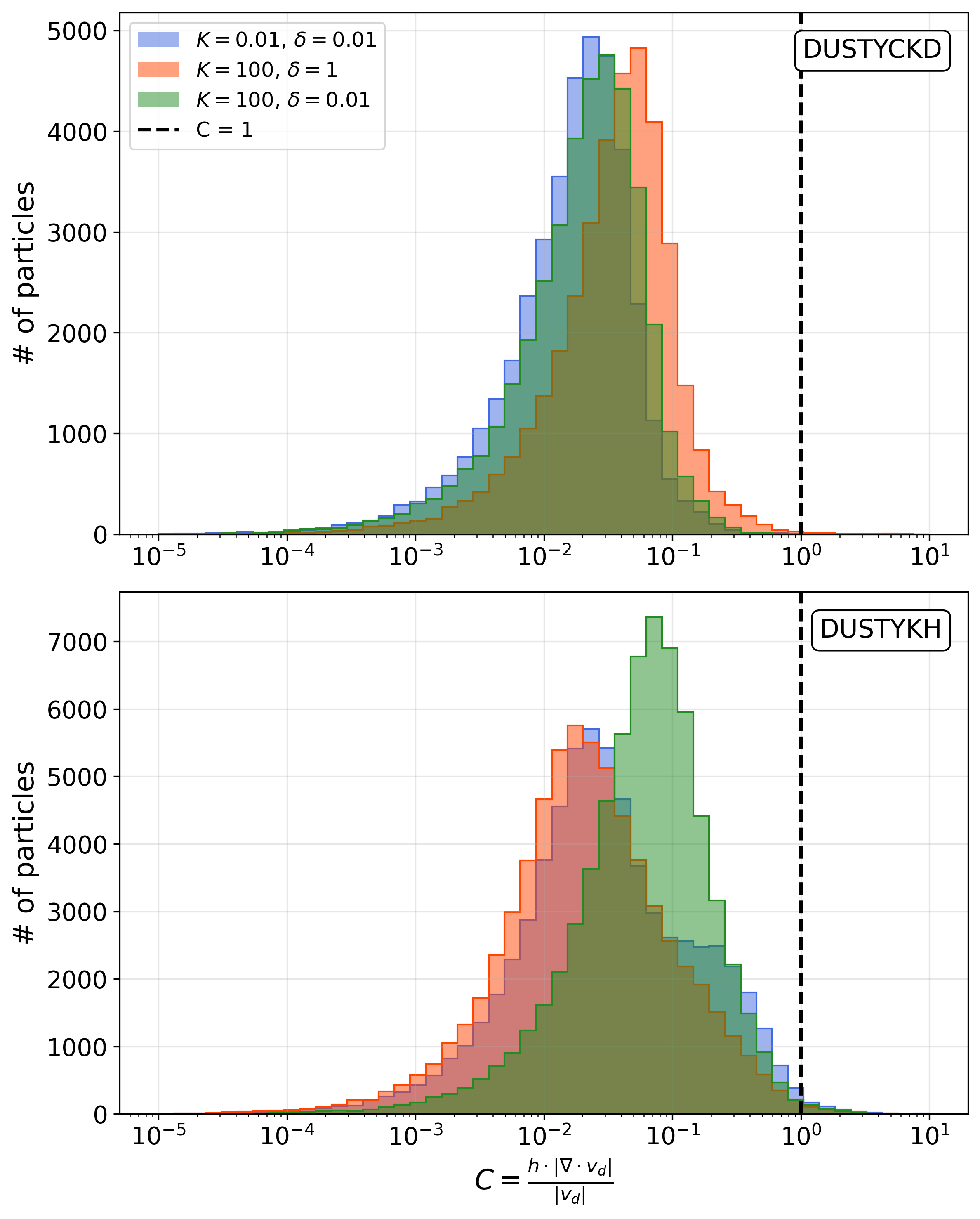}
    \caption{Distribution of the shock indicator defined in Eq. \ref{eq:shock_indicator} for the Cold Keplerian Disk (top panel) and the Kelvin-Helmholtz instability (bottom panel) simulations.}
    \label{fig:C_hist}
\end{figure}

One of the consequences of modeling dust as a pressureless fluid is its tendency, when the drag coefficient is sufficiently low, to form clumps of dust that cannot self-regulate and dissipate. This leads to large gradients in $\epsilon$, which will lead, eventually, to code crashes. In order to alleviate this tendency, we tried including some form of dust diffusion in the One Fluid model. Our first attempt was to tackle directly the evolution of $\rho_d$ by adding a Gradient Diffusion flux to its continuity equation. While this may certainly be a valid approach in Eulerian simulations, we found this approach not suitable for our SPH framework, as it would require adding a term in the continuity equation for the total mixture density, and hence a mass transfer term between SPH particles.\\
We thus resorted to a different approach, the inclusion of a "pressure-like" term in the dust momentum conservation equation, as described by \cite{Klahr2021} and \cite{Binkert2023}
\begin{equation}
    \frac{\partial}{\partial t}(\rho_d \vec{v}_d) + \nabla \cdot (\rho_d \vec{v}_d \vec{v}_d) = - \nabla \biggl(\frac{1}{3}\frac{D}{t_s}\rho_d\biggr) -\frac{\rho_d}{t_s}(\vec{v}_d - \vec{v}_g).
\end{equation}
Translating this approach in the One Fluid model equations, we obtain the following continuum equations
\begin{equation}
    \biggl(\frac{d\vec{v}}{dt}\biggr)_{diff} = -\frac{DK}{3\rho(1-\epsilon)^2}\nabla\epsilon = -\frac{2DKs}{3\rho}\nabla s,
\end{equation}
\begin{equation}
    \biggl(\frac{d\vec{\Delta v}}{dt}\biggr)_{diff} = -\frac{DK}{3\rho\epsilon(1-\epsilon)^2}\nabla\epsilon = -\frac{2sDK}{3\rho \epsilon}\nabla s,
\end{equation}
where we already provide the formulation for the dust fraction parametrization (Eq. \ref{eq:dust_frac_param}) that we will be using throughout the paper.\\
We propose the following suitable (momentum- and energy-conserving) SPH discretization for the previous equations
\begin{equation}
    \biggl(\frac{d\vec{v}_a}{dt}\biggr)_{diff} = -\sum_b m_b \frac{2\overline{D}_{ab}\overline{s}_{ab} K}{3\rho_a\rho_b} (s_b - s_a) \overline{\nabla \mathcal{W}_{ab}},
\end{equation}
\begin{equation}
    \biggl(\frac{d\vec{\Delta v}_a}{dt}\biggr)_{diff} = -\sum_b m_b \frac{2\overline{D}_{ab}\overline{s}_{ab} K}{3\rho_a\rho_b\overline{\epsilon}_{ab}} (s_b - s_a) \overline{\nabla \mathcal{W}_{ab}},
\end{equation}
\begin{equation}
    \biggl(\frac{du_a}{dt}\biggr)_{diff} = \sum_b m_b \frac{2\overline{D}_{ab}\overline{s}_{ab} K}{3\rho_a\rho_b(1-\overline{\epsilon}_{ab})} (s_b - s_a) (\vec{v}_a^d - \vec{v}_b^d) \cdot \hat{r}_{ab} \overline{F}_{ab},
\end{equation}
where the additional heating term was added to conserve energy, similarly to how Eq. \ref{dudtCAV} is introduced.\\
We only need to specify what diffusion coefficient $D$ to use. While some implementations of dust diffusion leave this choice to the user (as in \cite{Huang2022}), we want our implementation to be adaptive and only trigger when needed in order to make the dust density field smoother in case of large gradients. We tried different approaches, but what was found to have the best results in terms of stability was the following. We define a new variable, $\chi_a$, as the ratio between the dust fraction of particle $a$ and the mean of the dust fractions of all neighboring particles
\begin{equation}
    \chi_a = \frac{\epsilon_a}{\frac{1}{N_{ngb}}\sum_b \epsilon_b}.
\end{equation}
We want diffusion to act on particles for which $\chi_a$ departs from unity. A suitable definition for $D$ was found to be
\begin{equation}
    D_a = \frac{(\chi_a-1)^2}{\chi_a}.
    \label{eq:diff_coeff}
\end{equation}
In the SPH discretizations we use the symmetrical $\overline{D_{ab}}$ to ensure momentum and energy conservation.\\
This implementation could also be extended to describe a physical (or non-artificial) diffusion. This would require modeling the diffusion coefficient $D$ based on gas or dust properties, rather than on the numerical prescription Eq. \ref{eq:diff_coeff}.\\
In principle, defining a spatially varying diffusion coefficient could allow shocks to develop, which would then require a dedicated artificial-viscosity term. To test whether this actually occurs, we examined the ratio between the compression and advection timescales as a shock indicator. These timescales are defined as
\begin{equation}
    \tau_{\text{compr}} = \frac{1}{|\nabla \cdot \vec{v}_\text{d}|}, \quad \tau_{\text{adv}} = \frac{h}{|\vec{v}_\text{d}|}.
\end{equation}
In a shock, compression is much faster than advection alone would allow, and thus $\tau_{\text{compr}} < \tau_{\text{adv}}$. We thus define the shock indicator for dust 
\begin{equation}
\label{eq:shock_indicator}
    C = \frac{h \cdot |\nabla \cdot \vec{v}_{\text{d}}|}{|\vec{v}_\text{d}|},
\end{equation}
so that a shock happens in the dust when $C > 1$.\\
In Figure \ref{fig:C_hist} we show the histograms of the values for $C$ for the Cold Keplerian Disk and the Kelvin-Helmholtz instability simulations discussed, respectively, in Section \ref{sec:DUSTYCKD} and Section \ref{sec:DUSTYKH}. For all the drag coefficients $K$, and dust-to-gas ratios $\delta$, in simulations the vast majority of particles show values of $C < 1$, indicating the absence of dust shocks.
\\

\section{Tests}
In this section, we aim at showing both the accuracy and the stability of the implementation of the One-Fluid model in \texttt{OpenGadget3} shown in the previous section. We first show that the code is able to reproduce the expected results from standard dust dynamics tests: the DUSTYBOX, DUSTYWAVE, and DUSTYSHOCK test,s as well as a dusty Sedov-Taylor blast wave. For all these tests, we will not use our new dust diffusion model. Afterward, we test the stability of the code under more complex flows for various dust-to-gas ratios and drag coefficients. One of the major challenges in simulating dust dynamics is the tendency in low drag and high dust-to-gas ratio environments for dust to clump and create large gradients in dust fraction. To study these behaviors and find numerical solutions to make the code stable under these extreme circumstances, we simulate a Cold Keplerian Disk, a dusty protoplanetary disk, and a Kelvin-Helmholtz instability. These more complex flows will be simulated with dust diffusion turned on. All the values reported in the following section are in internal code units.\\

\subsection{DUSTYBOX}
\begin{figure}[h]
    \centering
    \includegraphics[width=1\linewidth]{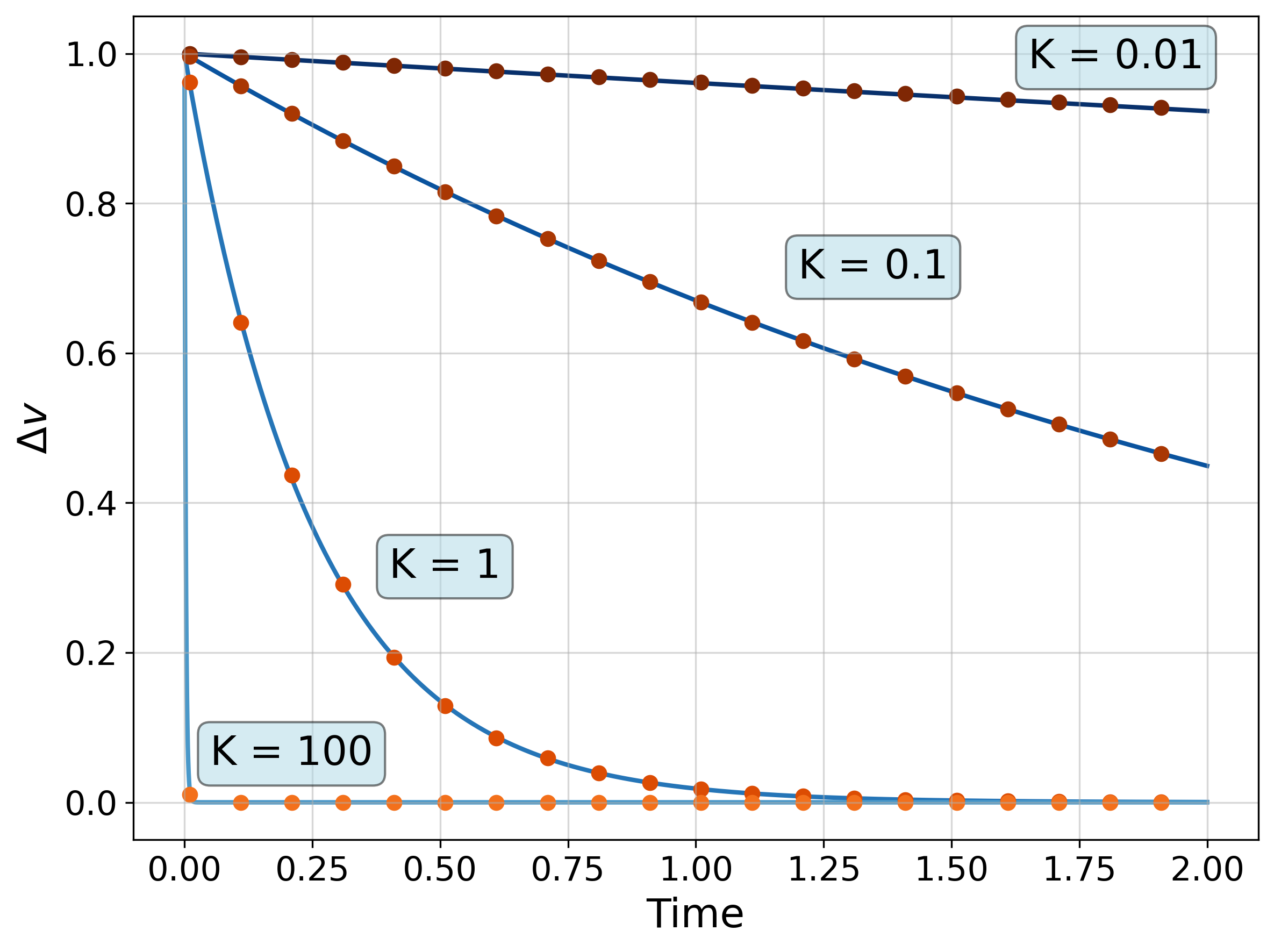}
    \caption{Result of the DUSTYBOX benchmark test. The continuous lines represent the analytical solution, while the dots are the simulation outputs. The results are shown for four different values of the drag coefficient $K$.}
    \label{fig:DUSTYBOX}
\end{figure}

The first benchmark test we perform is the so-called DUSTYBOX test. It consists of a mixture of gas and dust moving with respect to each other, but globally at rest. This is primarily used to test the integration of the drag term, since all other terms in the evolution equations are zero. The setup is extremely simple: 100 particles are equally spaced in one dimension, the density is set to $\rho_0 = 1$, and the dust fraction is set to $\epsilon_0 = 1/2$, meaning that there is an equal amount of dust and gas. The barycenter velocity is zero, $\vec{v} = 0$, so the mixture is at rest and the particles do not move in space. Additionally, the pressure is constant and set to $P_0 = 1$. \\
The only dynamics arises from the dust-gas velocity difference, which is initialized to $\vec{\Delta v}_0 = 1$. The expected behavior is for this initial velocity difference to decay exponentially, following the simple analytical solution
\begin{equation}
    \vec{\Delta v}(t) = \vec{\Delta v}_0 e^{-\frac{t}{t_s}}.
\end{equation}
The stronger the drag coefficient $K$, the smaller the stopping timescale $t_s$ and the faster the exponential decay.\\
In Figure \ref{fig:DUSTYBOX}, we show the results of our simulations for four different values of the drag coefficient. The simulation outputs show very good agreement with the expected behaviors. More quantitatively, we found the L1 norm to be always smaller than $< 0.1\%$. In addition, energy is also very well conserved: as the simulation evolves, the kinetic energy stored in the second term of the expression for the total energy in Eq. \ref{eq:total_energy} is converted into internal energy of the gas, via drag heating. The resulting energy is conserved up to differences of $10^{-7}$.

\subsection{DUSTYWAVE}
\begin{figure}
    \centering
    \includegraphics[width=1\linewidth]{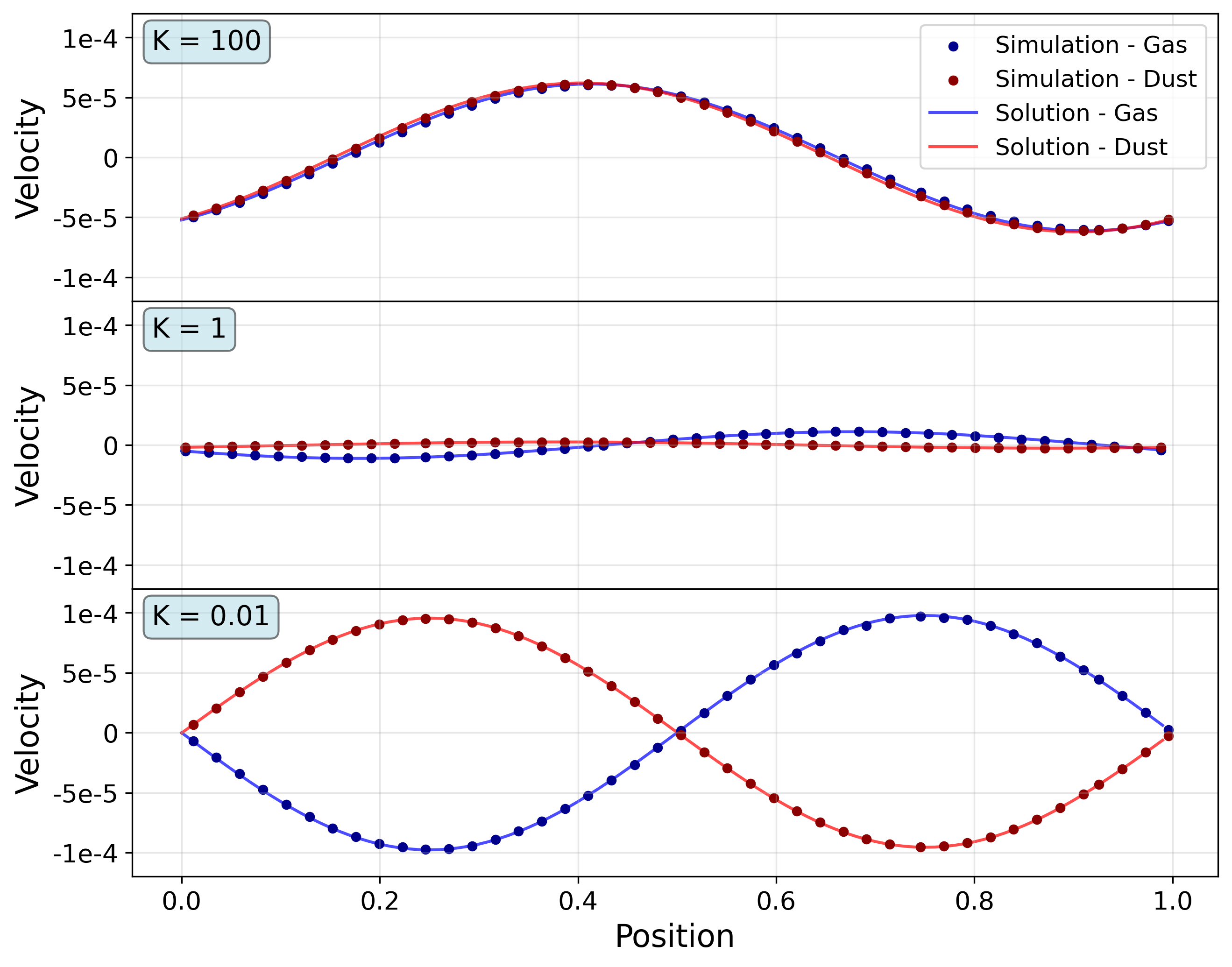}
    \caption{Results from the DUSTYWAVE benchmark test, for three different values of drag coefficient $K$. The continuous lines represent the analytical solution and the dots the output of the simulations. Gas and dust velocities are plotted separately.}
    \label{fig:DUSTYWAVE}
\end{figure}
The next benchmark test is a simple linear wave of a mixture of dust and gas. Since dust is a pressureless fluid, the motion is primarily driven by gas pressure, which makes the gas wave move across the domain. The dust is dragged by the gas according to the strength of the drag coefficient.\\
The wave is initialized by perturbing an equilibrium particle distribution. The particles span a periodic domain $x\in(0,1)$. The initial conditions for all variables are
\begin{align*}
     & \rho(x,0) = \rho_0 + A \sin(2\pi x), \\
     & \vec{v}(x,0) = A \sin(2\pi x),       \\
     & \epsilon(x,0) = 1/2,                 \\
     & \vec{\Delta v}(x,0) = 0,             \\
\end{align*}
where $A = 10^{-4}$ is the amplitude of the perturbation and $\rho_0 = 1$ is the background density value. The pressure is initialized with an isothermal equation of state, $P = \rho c_s^2$, with sound speed $c_s = 1$.\\
What makes the DUSTYWAVE a useful benchmark test is the availability of an analytical solution, derived by \cite{Laibe2011}.\\
The results of our simulations are shown in Figure \ref{fig:DUSTYWAVE}, for three values of drag coefficient. Depending on the strength of the drag coefficient, the dynamics of dust and gas can differ widely. We find very good agreement between simulations and analytical solutions, with $L_1$ norms of the order of $10^{-7}$. Also in this case energy is very well conserved. This is also achieved by artificial viscosity being active only in its conservative formulation, since the shock indicator here is very close to $0$.

\subsection{DUSTYSHOCK}
\begin{figure*}
    \centering
    \includegraphics[width=0.9\linewidth]{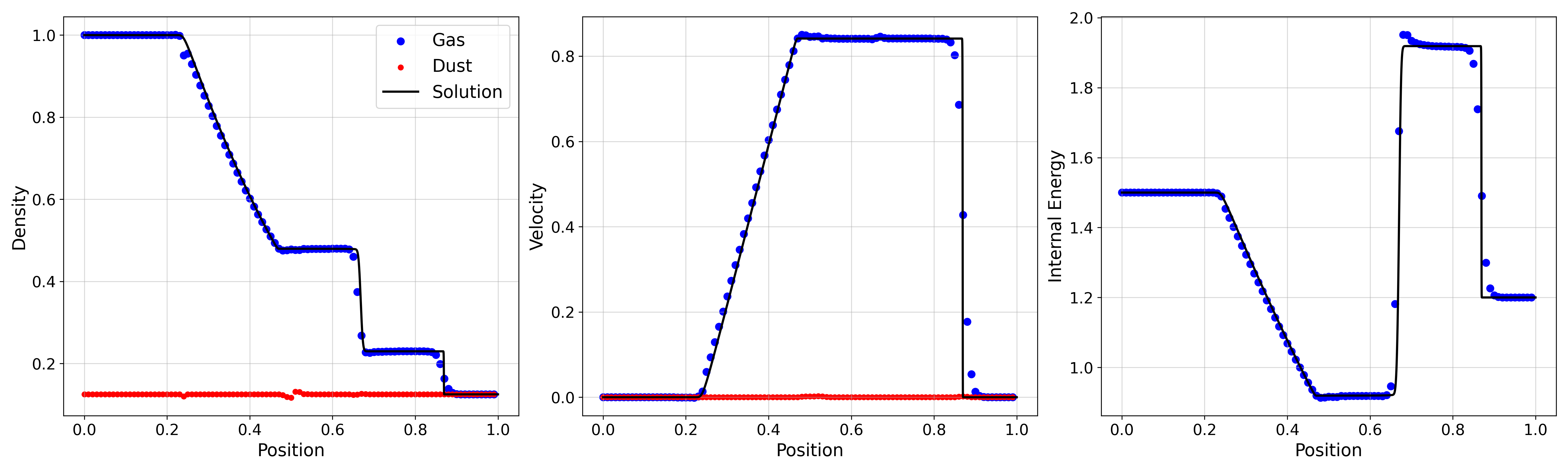}
    \caption{DUSTYSHOCK benchmark test for zero drag. The blue and red dots represent the output of the simulation for gas and dust, respectively, while the continuous line shows the solution for the gas evolution. As expected, while the gas undergoes the typical Sod shock tube evolution, the dust remains unperturbed.}
    \label{fig:DUSTYSHOCK_1D_K0}
\end{figure*}
\begin{figure*}
    \centering
    \includegraphics[width=0.9\linewidth]{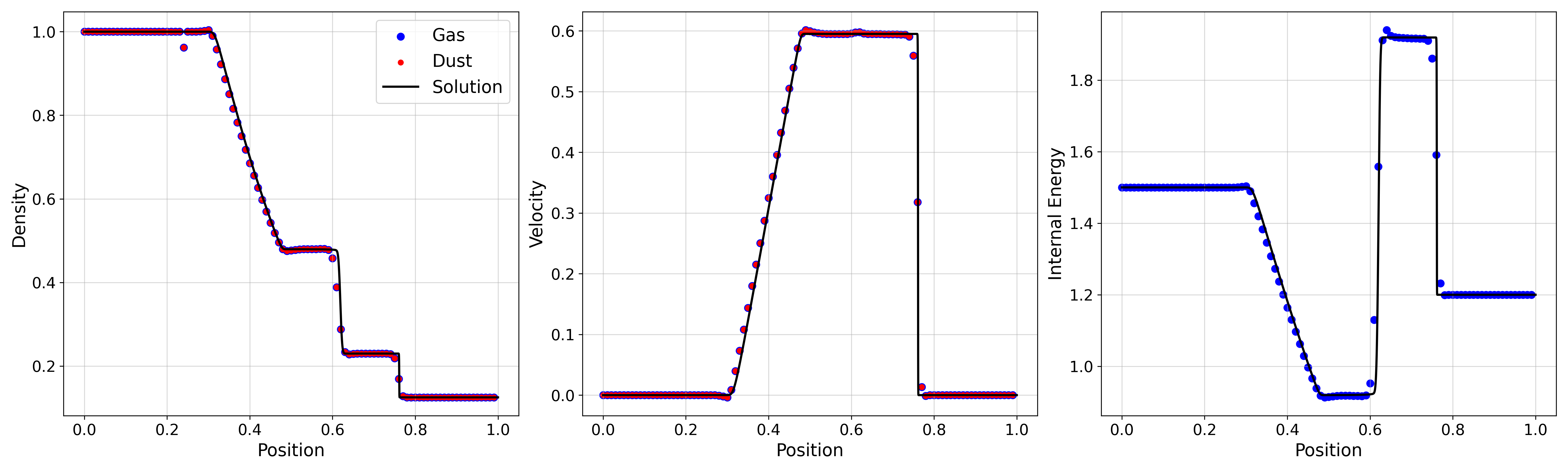}
    \caption{DUSTYSHOCK benchmark test for strong drag ($K = 10^6$). The blue and red dots represent the output of the simulation for gas and dust, respectively, while the continuous line shows the solution for the gas evolution. The evolution of the dust now closely follows the gas, and both evolve at a lower speed compared to a gas-only simulation, with the modified sound speed shown in Eq. \ref{eq:modified_sound_speed}.}
    \label{fig:DUSTYSHOCK_1D_Kinf}
\end{figure*}
The last suite of benchmark tests consists of a set of shocks. This is useful to test fully the implementation of the One-Fluid model, as shock environments require all of our artificial viscosity and conductivity terms. We explore two opposite scenarios: first, a weak drag regime, in which dust is expected to remain still as the shock propagates in the gas; secondly, a strong drag regime, where dust and gas move tightly coupled, and the shock propagates as a gas-only shock with modified sound speed. For both these regimes, we can compute a numerical solution using a Riemann Solver (here we chose an Exact Riemann Solver).\\
The DUSTYSHOCK test was already used in \citetalias{Laibe2014b} to test the One-Fluid model implementation in one dimension, but here we extend the test also to two dimensions.

\subsubsection{Weak coupling}
The gas is initialized as in a standard Sod shock tube test \citep{Sod1978}. The left and right states are
\begin{align*}
     & \rho_{L}^g = 1 \quad \rho_{R}^g = 0.125,   \\
     & \vec{v}_{L}^g = 0 \quad \vec{v}_{R}^g = 0, \\
     & P_L = 1 \quad P_R = 0.1.
\end{align*}
The dust, instead, is initialized as a constant background density at rest
\begin{align*}
     & \rho_L^d = 0.125 \quad \rho_R^d = 0.125,   \\
     & \vec{v}_{L}^d = 0 \quad \vec{v}_{R}^d = 0.
\end{align*}
The drag coefficient is set to $K=0$. The purpose of this test is to verify that the One-Fluid model implementation is able to evolve the gas shock without perturbing the dust background uniform density, which should remain unperturbed given the zero drag it feels from the gas.\\
The simulation is initialized with a total of $N=1375$ particles in a domain $x\in(-0.5,0.5)$. Boundary particles are placed outside each end of the domain to keep the particles confined. The internal energy is initialized and evolved following an ideal gas law, with constant $\gamma = 5/3$.\\
The results are shown in Figure \ref{fig:DUSTYSHOCK_1D_K0}. As expected, the gas undergoes the standard shock as if the dust were not present, and the dust, conversely, stays unperturbed. This excellent performance in keeping the dust still is to be attributed to the Non-Conservative artificial viscosity scheme, which is here fully active in the shocked regions of the fluid.\\

\subsubsection{Strong coupling}
\begin{figure*}
    \centering
    \includegraphics[width=0.9\linewidth]{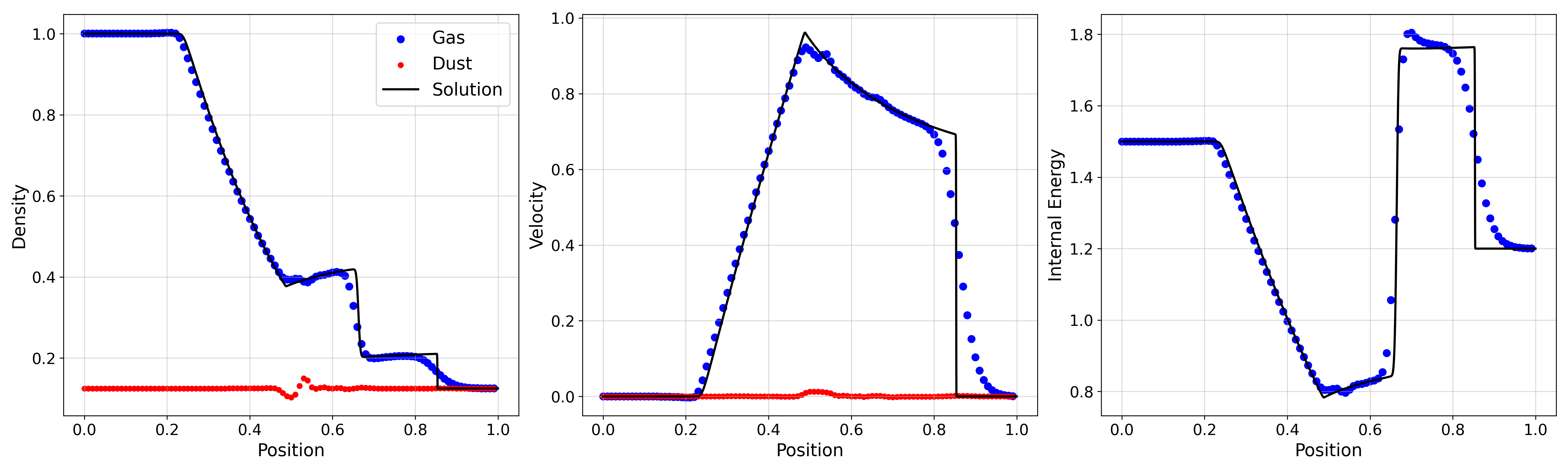}
    \caption{Same as in Figure \ref{fig:DUSTYSHOCK_1D_K0}, but now in two dimensions. The solution was obtained with a one-dimensional Exact Riemann Solver with an additional geometrical source term as described in Eq. \ref{eq:geom_source_term}. The dust now shows some additional wiggles compared to the one-dimensional case, which can be attributed to the lower effective resolution of this higher-dimensional test.}
    \label{fig:DUSTYSHOCK_2D_K0}
\end{figure*}

\begin{figure*}
    \centering
    \includegraphics[width=0.9\linewidth]{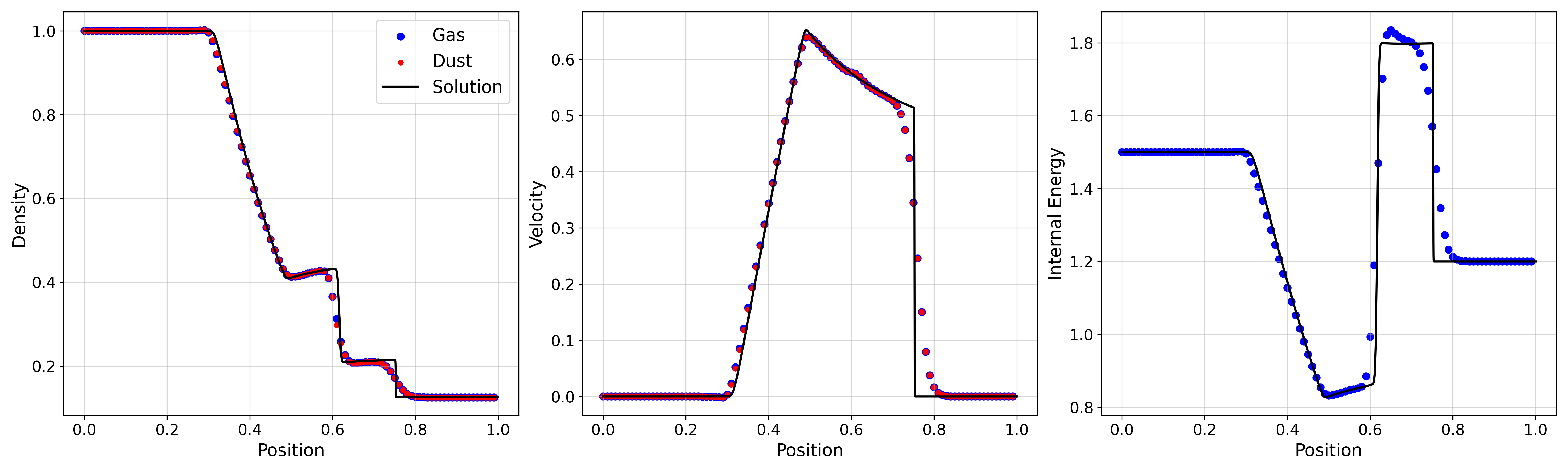}
    \caption{Same as in Figure \ref{fig:DUSTYSHOCK_1D_Kinf}, but now in two dimensions. The solution was obtained with a one-dimensional Exact Riemann Solver with an additional geometrical source term as described in Eq. \ref{eq:geom_source_term}. }
    \label{fig:DUSTYSHOCK_2D_Kinf}
\end{figure*}
We also explore the opposite scenario, a Sod shock tube test, where dust is strongly coupled to the gas. While the gas initial left and right states are identical to the ones used for the weak coupling case, the dust is now initialized using the same density distribution as the gas, as opposed to the constant background used previously
\begin{align*}
     & \rho_L^d = 1 \quad \rho_R^d = 0.125,       \\
     & \vec{v}_{L}^d = 0 \quad \vec{v}_{R}^d = 0.
\end{align*}
The drag coefficient is set to $K = 10^{6}$ and a total of $N = 1700$ particles are employed.\\
The expected behavior for this shock is to evolve as a standard Sod shock tube, but with a modified sound speed of
\begin{equation}
    \tilde{c_s} = \frac{c_s}{\sqrt{1-\epsilon}}.
    \label{eq:modified_sound_speed}
\end{equation}
As can be shown in Fig. \ref{fig:DUSTYSHOCK_1D_Kinf}, the result from the simulation follows closely the expected solution, and dust and gas remain tightly coupled.

\subsubsection{Two-dimensional shocks}
In order to test the One-Fluid model implementation in two dimensions, we employed a strategy discussed in \cite{Toro2009} to turn one-dimensional shocks into two-dimensional radially symmetric shocks. In particular, \cite{Toro2009} shows that the radial evolution of a shock in multiple dimensions can be replicated in one-dimensional simulations by adding a geometrical source term to the Euler equations
\begin{equation}
    \partial_t\vec{U} + \partial_r\vec{F}(\vec{U}) = \vec{S}(\vec{U}),
    \label{eq:geom_source_term}
\end{equation}
where $\vec{U}$ is the vector of conserved variables, $\vec{F}(\vec{U})$ are the corresponding fluxes and the geometrical source term $\vec{S}(\vec{U})$ is defined as
\begin{equation}
    \vec{S}(\vec{U}) = -\frac{\alpha}{r}
    \begin{bmatrix}
        \rho v   \\
        \rho v^2 \\
        v(E+p)
    \end{bmatrix}.
\end{equation}
By varying the $\alpha$ parameter we obtain a cylindrical ($\alpha = 1$) or a spherical ($\alpha = 2$) symmetry. With this procedure, we can obtain the expected result for two-dimensional simulations using a one-dimensional Riemann solver.\\
The initial conditions for the shocks in two dimensions are essentially the same as in the one-dimensional shock. The particles are arranged in a hexagonal closed pack lattice in a periodic domain $x\in(-1,1)$, $y\in(-1,1)$. Particles inside a sphere centered around $0$ and of radius $R = 0.5$ have the "left" state initial conditions, while the particles outside have the "right" state initial conditions. The results for the two-dimensional shocks for both the weak and strong drag cases are shown in Fig. \ref{fig:DUSTYSHOCK_2D_K0} and Fig. \ref{fig:DUSTYSHOCK_2D_Kinf}. The performances are slightly worse compared to the one-dimensional case, which can be explained by a lower effective resolution of the multidimensional cases, but the shocks all evolve as expected.

\subsection{Dusty Sedov-Taylor blast wave}
\begin{figure*}
    \centering
    \includegraphics[width=0.9\linewidth]{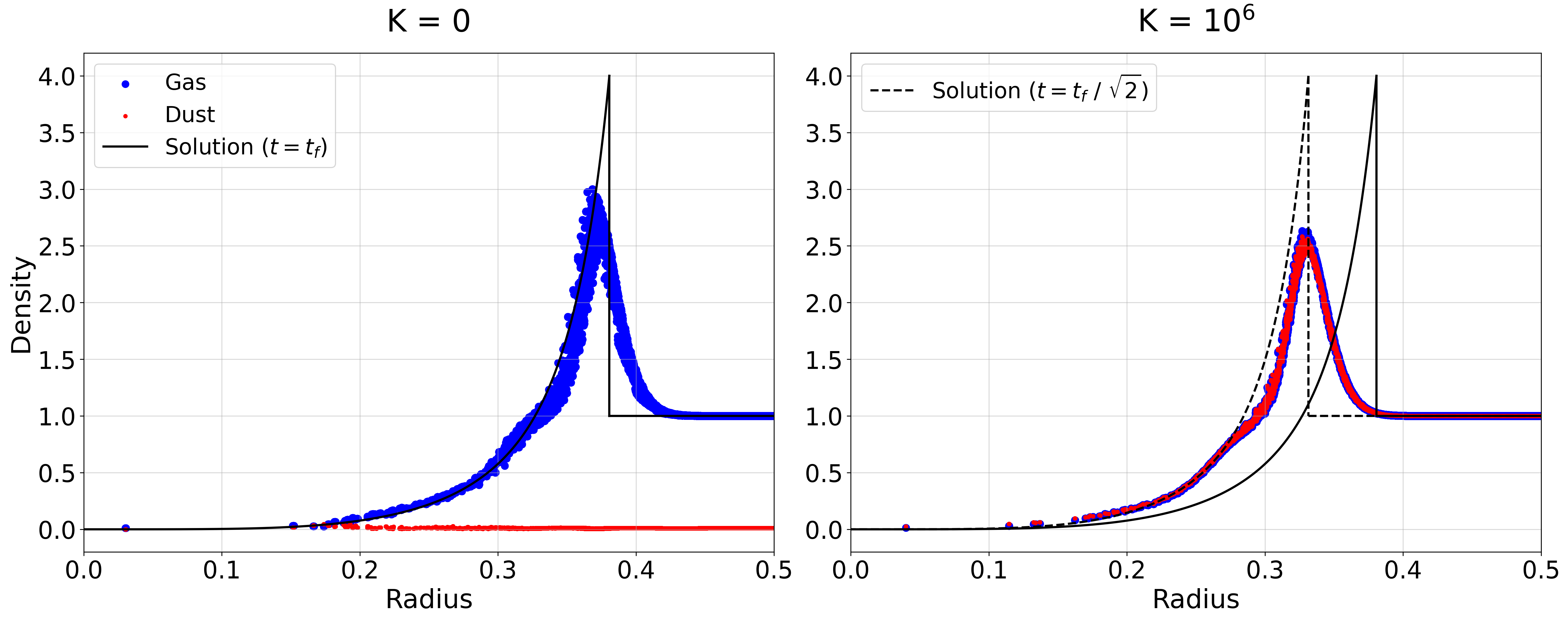}
    \caption{Dusty Sedov-Taylor blast wave for a dust and gas mixture in two regimes, $K=0$ (left) and $K=10^6$ (right). The radial density profiles are shown as blue dots for gas and red dots for dust, and the self-similar solution at $t=t_\text{f}$ is plotted as a continuous black line. The simulation correctly keeps dust at rest while the gas evolves following the self-similar solution, for uncoupled dust. For strongly coupled gas and dust, the mixture follows the self-similar solution evaluated at a previous time, consistent with the modified sound speed in Eq. \ref{eq:modified_sound_speed}, shown as a dashed black line.}
    \label{fig:DUSTYSEDOV}
\end{figure*}
The Sedov-Taylor blast wave is a strong, radially symmetric shock wave, first introduced by \cite{Sedov1946, Sedov1959}. It is widely used in astrophysics to model the evolution of supernova blast waves \citep{Steinwandel2020} and to benchmark the shock-capturing capabilities and limitations of hydrodynamical simulation codes \citep{Springel2002, Rosswog2007, Groth2023}.\\
For gas-only simulations, a self-similar solution exists and can be used to verify the performance of numerical schemes. Although no such solution is known for dust–gas mixtures, in the limit of strong drag and high dust-to-gas ratios, the evolution is expected to follow the gas-only solution but at a modified sound speed, as in the \textit{DUSTYSHOCK} tests.\\
In the context of dust-gas mixtures, \cite{Laibe2012a} simulated a Sedov-Taylor blast wave using a two-fluid model for a moderate drag coefficient of $K=1$, since a high drag simulation would have been too computationally expensive, and a zero-drag simulation would have been trivial in the two-fluid model. Here we will simulate the Sedov-Taylor blast wave for a mixture of dust and gas in two cases, mirroring the ones used in the DUSTSHOCK tests: (i) a zero-drag simulation with dust-to-gas ratio of $0.01$ and (ii) a high-drag ($K = 10^6$) simulation with dust-to-gas ratio of $1$.\\
We initialize $N = 64^3$ particles in a regular grid $x,y,z \in (-0.5,0.5)$ with periodic boundary conditions, uniform gas density $\rho_g = 1$ and vanishingly low pressure $P=10^{-6}$, except for the particles closest to the center which are initialized with a large internal energy $U = 10$. Consequently, a shock with Mach number $ M\sim 2\cdot 10^{4}$ propagates through the domain. The simulation is evolved until $t_\text{f} = 0.02$\\
In Figure \ref{fig:DUSTYSEDOV} we show the radial density profiles obtained from the simulations for both regimes, zero and high drag, together with the self-similar solution. For $ K=0$ (left), the gas undergoes the Sedov-Taylor blast wave and approaches the expected self-similar solution, while the dust remains largely at rest. For high drag (right), the dust and gas stay tightly coupled, and the mixture can be described by the self-similar solution evaluated at an earlier time, consistent with the modified sound speed given in \ref{eq:modified_sound_speed}.

\subsection{Dusty Cold Keplerian Disk}
\label{sec:DUSTYCKD}
\begin{figure*}[]
    \centering
    \includegraphics[width=0.9\linewidth]{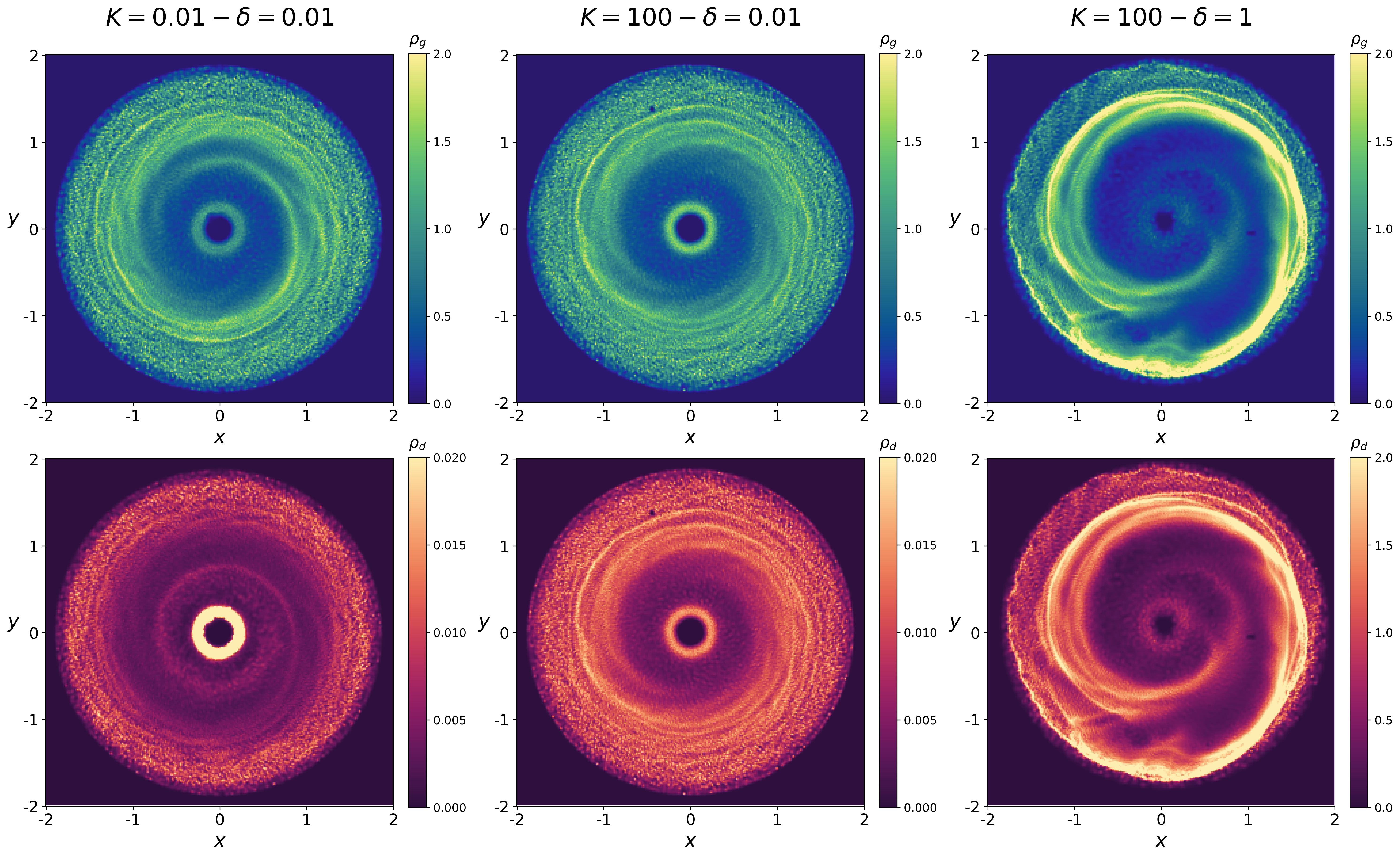}
    \caption{Dusty Cold Keplerian Disk simulations for three different sets of parameters. Gas density is shown in the top panels while dust density is shown in the bottom panels.}
    \label{fig:DUSTYCKD}
\end{figure*}
The Cold Keplerian Disk is a standard hydrodynamical test used to study the stability of numerical frameworks \citep{Cartwright2009, Hopkins2015, Groth2023}. SPH has already been shown to struggle in keeping a Cold Keplerian Disk stable for various orbits; thus, we are not interested in studying its stability. Instead, we will focus on how dust clumping behaves under various dust-to-gas ratios and drag coefficients and how the dust pressure model introduced in Section 2.7 can help keep the code stable.\\
The initial condition for the Cold Keplerian Disk is taken, for instance, from \cite{Groth2023}. The gas surface density is given by
\begin{equation}
    \Sigma_g = 0.01 + \begin{cases}
        \biggl(R / 0.5\biggr)^3 \quad        & \text{if  } R \leq 0.5,        \\
        1 \quad                              & \text{if  } 0.5 \leq R \leq 2, \\
        \biggl(1+(R-2)/0.1\biggr)^{-3} \quad & \text{if  } R > 2.
    \end{cases}
\end{equation}
\begin{figure}
    \centering
    \includegraphics[width=0.9\linewidth]{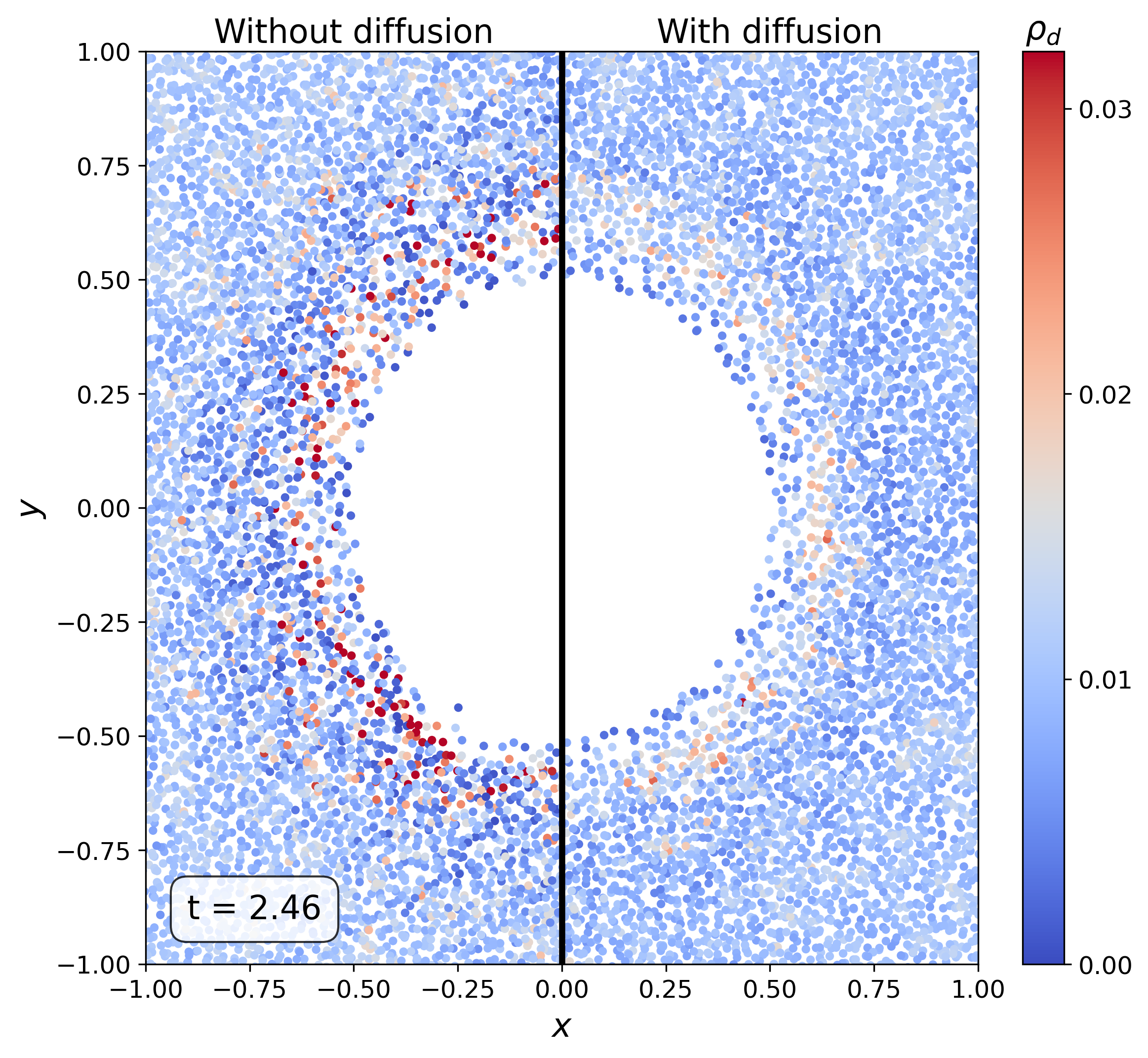}
    \caption{Zoom-in in the central region of the Cold Keplerian Disk simulation with weak drag ($K = 0.01$) and low dust-to-gas ratio ($\delta = 0.01$). The left half shows the last snapshot of a run without diffusion. The lack of a restoring force causes strong clumping of dust density, which leads to a code crash very early into the simulation. The addition of diffusion, as shown in the right half of the figure, dissipates these clumps and makes the code more stable.}
    \label{fig:DUSTYCKD_check_diffusion}
\end{figure}
Particles are initially distributed in a hexagonal close-packed grid in a domain $x\in(-2,2)$, $y\in(-2,2)$, and then removed from $R < 0.5$ and $R > 2$ until the desired density profile is reached. The effective resolution of the initial grid is $256^2$. The velocity is set to strictly Keplerian, and the pressure is set at a vanishingly low value of $P = 10^{-6}$. In adding the dust we simply increase the density uniformly by a factor $(1+\delta)$, where $\delta$ is the dust-to-gas ratio. The dust fraction is set accordingly as a uniform $\epsilon = \delta / (1+\delta)$. Dust is also initialized at a Keplerian velocity, hence, the initial difference velocity vector is $\vec{\Delta v} = 0$.\\
Such initial conditions should remain at equilibrium indefinitely, but in SPH the artificial viscosity eventually causes the disk to break up after roughly 5-10 orbits \citep{Hopkins2015}. We conduct a series of simulations varying the initial dust-to-gas ratio and the drag coefficient $K$.\\
In Figure \ref{fig:DUSTYCKD_check_diffusion} we show two simulations, both with $\delta = 0.01$ and $K = 0.01$, the only difference being the use of dust pressure as a diffusive term, as described in Section 2.8. The left half of the figure, corresponding to a simulation without any diffusion, shows severe clumping in dust density, which eventually causes a crash in the code. The snapshot shown here is the last available from the simulation. On the right half is shown a simulation where diffusion is switched on, at the same snapshot. As shown in the figure, the dust density field is less clumped, resulting in smoother gradients and a more stable simulation.
\\
Figure \ref{fig:DUSTYCKD} presents the results of three distinct simulation configurations at time $t=120$, corresponding to roughly 20 orbits at $R=1$. As anticipated, the disk becomes unstable over time in all cases due to the artificial viscosity inherent in SPH. The left panels depict a simulation with weak drag ($K = 0.01$) and a low dust-to-gas ratio ($\delta = 0.01$). In this scenario, the dust gradually drifts inward as it loses angular momentum, but its influence on the gas dynamics is negligible due to the small dust-to-gas ratio. The middle panels illustrate a simulation with strong drag ($K = 100$) and the same low dust-to-gas ratio ($\delta = 0.01$). Here, the dust closely follows the motion of the gas, as expected under strong coupling conditions. Finally, the right panels show a simulation with strong drag ($K = 100$) and a high dust-to-gas ratio ($\delta = 1$). In this case, the disk undergoes significant disruption, as the pressure gradient force of the gas is insufficient to counteract the influences from the dust dynamics. The dust concentrates in spiral structures that actively drag the gas along, resulting in a highly coupled system where the gas motion is strongly influenced by the dust.\\
While the presence of dust diffusion helped stabilize the code under various conditions, especially for weak drag regimes, the code could not remain stable for the most extreme conditions of weak and intermediate drag ($K = 0.01$ and $K=1$) for a dust-to-gas ratio of unity ($\delta = 1$). Although our code aims at being as general-purpose as possible, such extreme conditions are rarely encountered in common astrophysical applications and would require specialized techniques to be simulated.

\subsection{Dusty Protoplanetary Disk}
\begin{figure*}
    \sidecaption
    \includegraphics[width=0.6\linewidth]{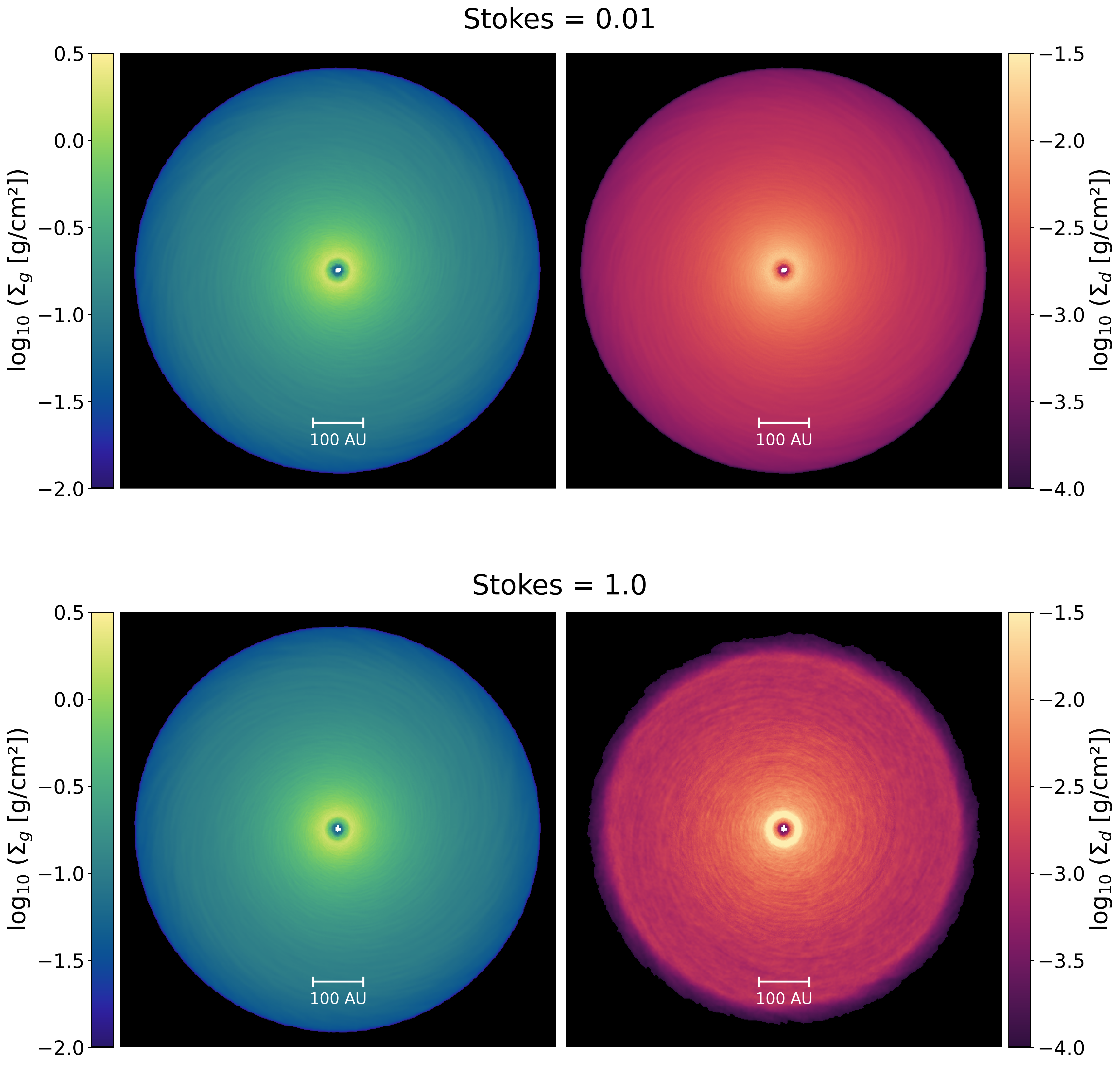}
    \caption{Gas (left panels) and dust (right panels) density for a dusty protoplanetary disk for two values of Stokes number at $t=20~kyr$. For tightly coupled grains ($St = 0.01$, top row), the gas and dust density remains at equilibrium, while for weakly coupled grains ($St = 1$, bottom row), the dust undergoes both clumping and radial drift toward the pressure maximum.}
    \label{fig:DUSTYPPD}
\end{figure*}
\begin{figure}
    \centering
    \includegraphics[width=0.8\linewidth]{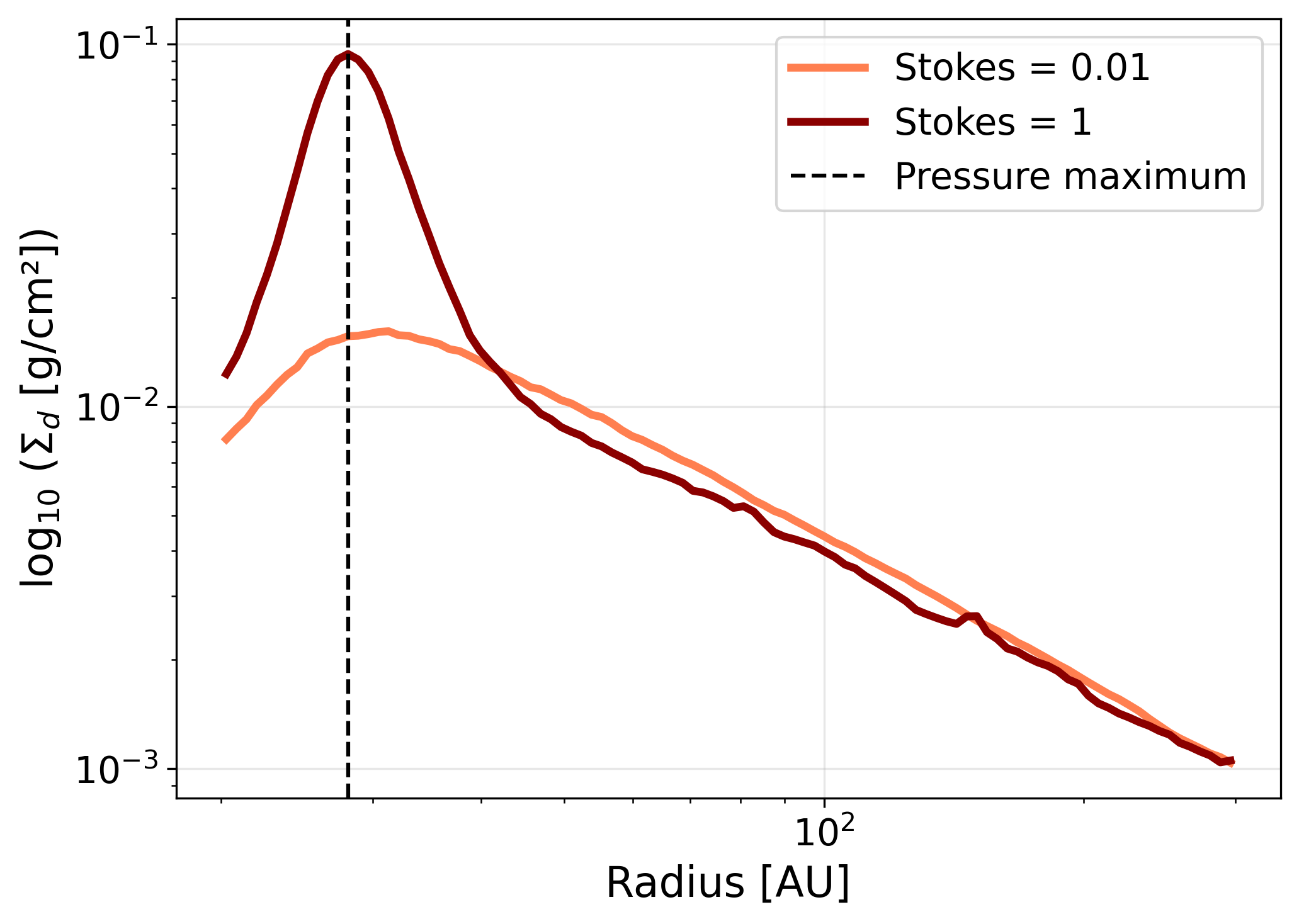}
    \caption{Dust radial density profile for both runs with $St=0.01$ and $St=1$. The coupled grains keep their original density profile, driven by the strong drag force, which forces them to follow the dynamics of the gas. The weakly coupled grains, on the other hand, undergo a much stronger radial drift toward the position of the pressure maximum.}
    \label{fig:DUSTYPPD_profiles}
\end{figure}
\begin{figure*}
    \centering
    \includegraphics[width=0.9\linewidth]{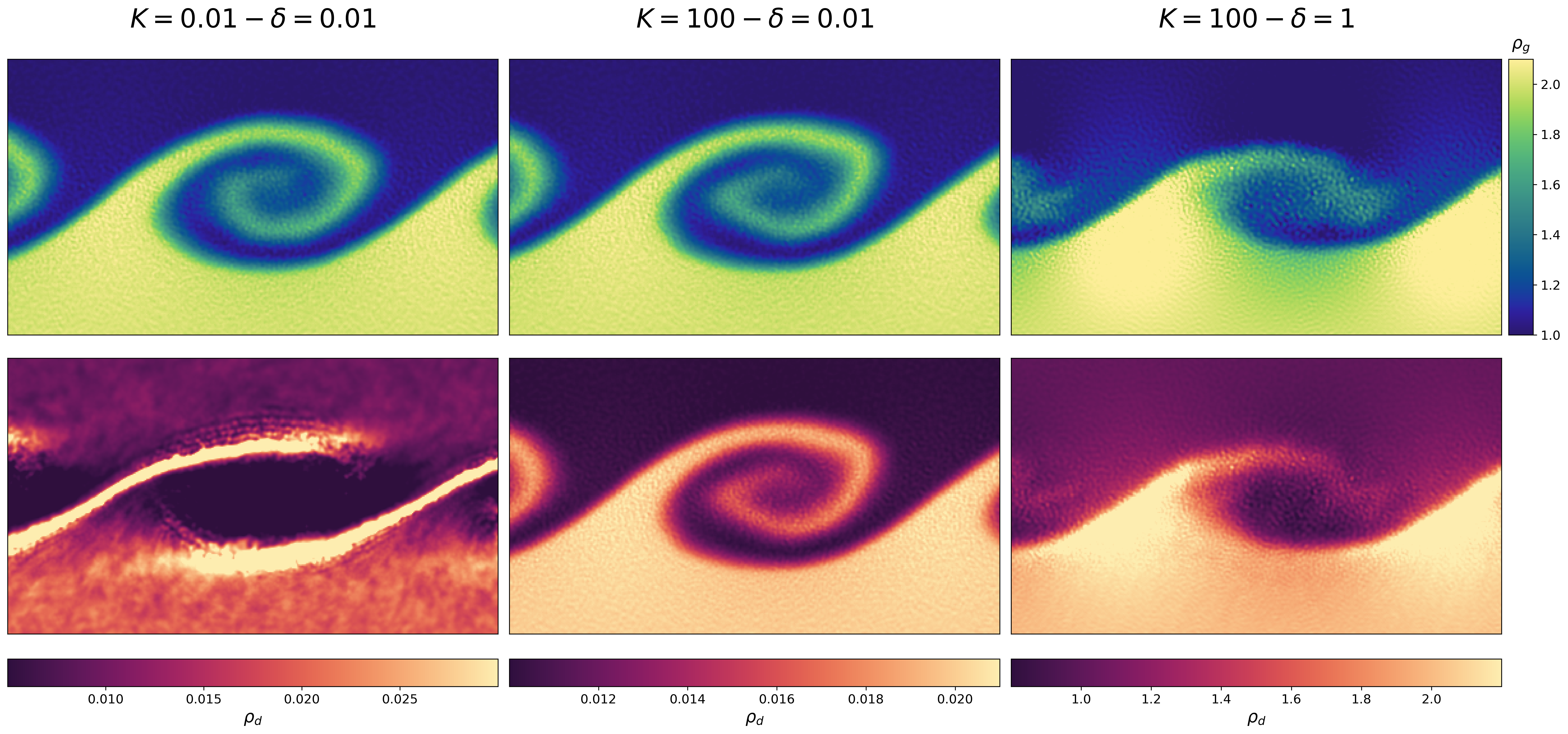}
    \caption{Dusty Kelvin-Helmholtz instability simulations for three different sets of parameters. Gas density is shown in the top panels while dust density is shown in the bottom panels.}
    \label{fig:DUSTYKH}
\end{figure*}
\begin{figure*}
    \centering
    \includegraphics[width=0.9\linewidth]{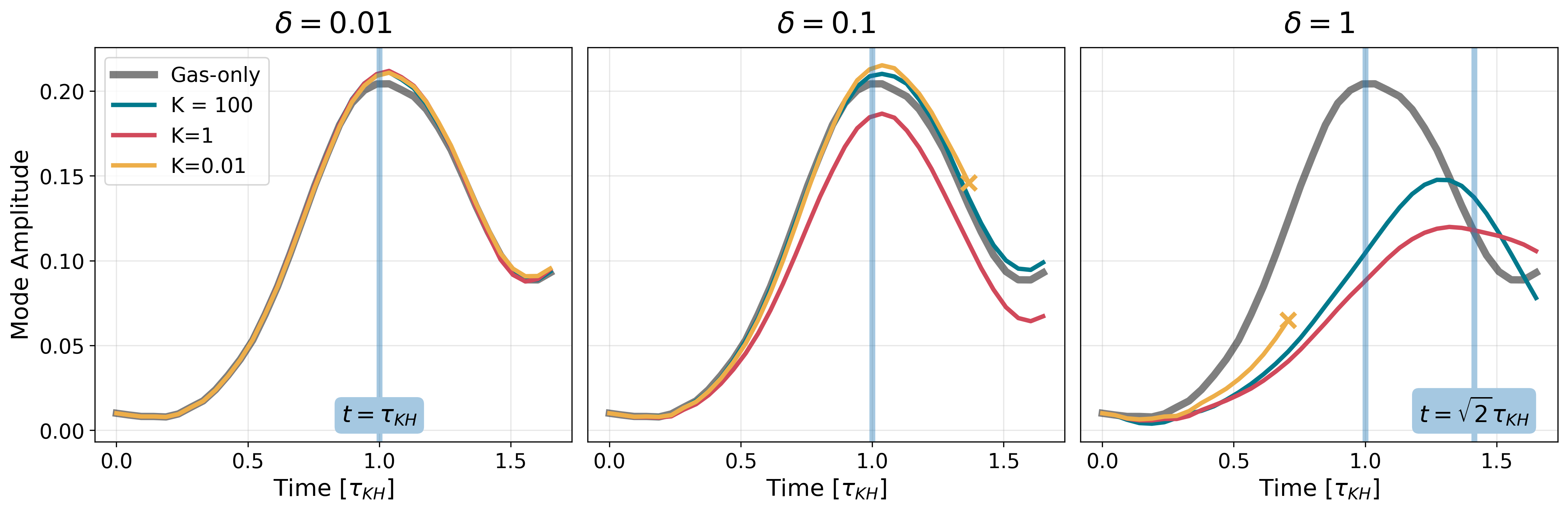}
    \caption{Time evolution of the Kelvin-Helmholtz instability mode amplitude for three dust-to-gas ratios, $\delta = (0.01,0.1,1)$, and three drag coefficients, $K=(0.01, 1, 100)$. Time is shown in units of the growth timescale $\tau_{KH}$. }
    \label{fig:DUSTYKH_growth}
\end{figure*}
The Cold Keplerian Disk is a useful and challenging hydrodynamical test to investigate the stability of our implementation. However, a more standard and physically motivated setup is a model of an actual protoplanetary disk. Our configuration follows previous SPH studies of protoplanetary disks \citep{Laibe2012b, Vericel2021}, but restricted here to two dimensions.\\
The surface density distribution follows a power law with a smooth inner edge
\begin{equation}
    \label{eq:dens_ppd}
    \Sigma(R) = \Sigma_0R^{-p}\left(1 - \sqrt{\frac{R_{\text{in}}}{R}}\right),
\end{equation}
with $p = 3/2$ and $R_{\text{in}} = 10~\text{au}$, and $\Sigma_0$ chosen so that the total mass of the disk is $M_{\text{disk}} = 0.01 M_{\odot}$. To generate a particle distribution with this density profile, we first use inverse sampling of the cumulative distribution function to generate values for the radii, and random values for the $\phi$ coordinate. To reduce the statistical noise of the resulting point distribution, we apply the Centroidal Voronoi Tessellation (CVT) technique described in \citet{Du1999}. The initial particle set is used to construct the corresponding Voronoi tessellation with the \texttt{qvoronoi} command from the \texttt{Qhull} library \citep{Barber1996}. We then compute the mass centroid of each Voronoi cell as
\begin{equation}
    \vec{x}_{j,i+1} = \frac{\int_{V_{j,i}} \vec{x}\Sigma(\vec{x})dV}{\int_{V_{j,i}}\Sigma(\vec{x})dV},
\end{equation}
where $\Sigma$ is the target density profile in Eq. \ref{eq:dens_ppd}, the $j$ index spans over the points in the distribution and the $i$ denotes the iteration step. The new point distribution $\vec{x}_{j,i+1}$ is then used to draw a new Voronoi tessellation, and the procedure is iterated until convergence. The resulting point distribution is much smoother than the initial one obtained from inverse sampling.\\
The CVT was used to generate the positions of $N=10^6$ points between $R_{\text{in}}=10~\text{au}$ and $R_{\text{out}}=400~\text{au}$. The dust-to-gas ratio is set to 0.01, and both gas and dust follow the same initial density profile. The disk is locally isothermal, with the sound speed following the power-law
\begin{equation}
    c_s = c_{s,0} R^{-q/2},
\end{equation}
with the index $q=3/4$ and the sound speed at radius $R=1~\text{au}$ equal to the aspect ratio: $c_{s,0} = h = 0.0281$, as in \cite{Vericel2021}.\\
The azimuthal velocity is set to exactly Keplerian for dust, and slightly sub-Keplerian for the gas
\begin{equation}
    v_{g,\phi} = v_K\sqrt{1-h^2(p+q)},
\end{equation}
where $v_K$ is the Keplerian azimuthal velocity.\\
The dust drag coefficient is set through its Stokes number, and two values are used: $St = 0.01$ and $St = 1$, to study the different evolution of tightly and weakly coupled dust grains.\\
These initial conditions are evolved until $t_\text{max}=20~kyr$ under the influence of the gravitational potential of the central star of mass $M_* = 1~M_{\odot}$. During the evolution, particles that cross the inner radius $R_{\text{in}}$ are removed from the simulation.\\
The gas and dust density maps for both $\mathrm{St}=0.01$ and $\mathrm{St}=1$ at $t=t_{\text{max}}$ are shown in Figure \ref{fig:DUSTYPPD}. For tightly coupled grains, both the gas and dust density profiles remain stable for long timescales. The pressure support of the gas and the drag force acting on the dust make this test much more stable than the Cold Keplerian Disk counterpart. For $St=1$, the low dust content is not enough to modify the gas dynamics. However, dust undergoes more significant clumping and radial drift. In Figure \ref{fig:DUSTYPPD_profiles} we compare the radial dust density profiles for the two runs, and it is evident that the $St=1$ run favors radial drift toward the pressure maximum created by the inner radius tapering in Eq. \ref{eq:dens_ppd}.

\subsection{Dusty Kelvin-Helmholtz Instability}
\label{sec:DUSTYKH}
To explore the capability of the One Fluid model in simulating the dynamics of dust in turbulent motion, we simulate a Kelvin-Helmholtz instability for various choices of dust-to-gas ratios and drag coefficients. As before, we add the dust to a gas-particle setup by uniformly multiplying the density by (1+$\delta$). We use the same initial conditions as, for instance, \cite{McNally2012}, \cite{Hopkins2015} or \cite{Groth2023}: a periodic box in two dimensions is filled with two fluids of density $\rho_1 = 1$ and $\rho_2 = 2$ with opposite velocities $\vec{v}_1 = 0.5\hat{x}$ and $\vec{v}_2 = -0.5\hat{x}$ and a small perturbation in the $\hat{y}$ direction to trigger the instability. We expect the instability to grow with a timescale $\tau_{KH} = \frac{\lambda}{v_{x,1} - v_{x,2}}\frac{\rho_1+\rho_2}{\sqrt{\rho_1 \rho_2}}$. In Figure \ref{fig:DUSTYKH} we show both the gas and dust densities at a time $t \simeq 1.2 \tau_{KH}$ for the same dust-to-gas ratios and drag coefficients that were shown in Figure \ref{fig:DUSTYCKD} for the Cold Keplerian Disk.

In the left panels, corresponding to low dust-to-gas ratio and weak drag, gas is not affected by the presence of dust, while the dust feels just enough drag to cluster along the edges of the vortex without tracing the gas inside. In the middle panels (low dust-to-gas ratio and high drag), the small amount of dust is stirred so effectively by the gas that it closely follows the evolution of a gas-only simulation. For high drag and high dust-to-gas ratio (right panels), the growth of the instability is slowed and dampened considerably, as gas has to drag more material upward in order to develop its turbulent motion. Both the clustering of dust around the edges of the vortices for low drag and the slowing down of the growth rate for large dust-to-gas ratios were already reported by \cite{Hendrix2014}, confirming the equivalence of our SPH implementation with their Eulerian approach.\\
To explore in more detail the growth of the instability for various parameters, we compute the mode amplitude $M$ using discrete convolutions, as described by \cite{McNally2012}. The evolution of the mode amplitude is shown in Figure \ref{fig:DUSTYKH_growth} for all dust-to-gas ratios and drag coefficients. In the left panel ($\delta = 0.01$), all curves closely follow the evolution of a gas-only simulation, with the mode amplitude peaking, as expected, at time $t = \tau_{KH}$. Variations in mode amplitude between simulations with different drag coefficients start appearing for $\delta = 0.1$ (middle panel). In particular, the growth for K = 1 is damped, similar to what we observed for the dusty wave at intermediate drag regimes (i.e., where both gas and dust waves were significantly damped). For large dust-to-gas ratio ($\delta = 1$, right panel) the evolution is very different: the growth timescale increases and the amplitude of the growth gets dampened. Considering that the increased dust-to-gas ratio reduces the effective sound speed of the medium, as described in Eq. \ref{eq:modified_sound_speed}, a crude approximation for the new growth timescale is $\tau_{KH}' = \sqrt{2}\tau_{KH}$.\\
Note that the weak drag ($K=0.01$) simulations for $\delta = 0.1$ and $\delta =1$ did not reach the end of the simulation, as for the Cold-Keplerian Disks, because of too large gradients of dust fraction.

\section{Conclusions}
We have described the implementation of the One-Fluid model in \texttt{OpenGadget3}, built upon the original formulation and implementation by \cite{Laibe2014a} and \cite{Laibe2014b}. Improvements from the original implementations are:
\begin{itemize}
    \item A correct version of the equation for $\frac{d\vec{\Delta v}}{dt}$, both in the continuum and in its SPH-discretized formulation. The continuum version is presented in two forms: one more compact and one more convenient to discretize, for the energy conservation arguments presented in Section 3. Many other possible forms of this equation can be derived, such as the one presented in \cite{Lebreuilly2019}.
    \item A simple time-dependent extension of artificial viscosity and artificial conductivity, already present in \texttt{OpenGadget3} and adapted for a dusty fluid.
    \item A unified artificial viscosity framework that uses the shock indicator $R_a$ to automatically interpolate between the conservative and non-conservative formulations presented in \cite{Laibe2014b}, capitalizing on the benefits from both methods.
    \item A novel artificial diffusion model for dust, built upon the "pressure-like" term introduced by \cite{Klahr2021}, with an adaptive diffusion coefficient.
\end{itemize}
The implementation was benchmarked against standard dust dynamics tests such as DUSTYBOX, DUSTYWAVE, and DUSTYSHOCK, the last of which was performed in one and two dimensions. Additionally, we further tested the shock-capturing properties of our implementation against a dusty Sedov-Taylor blast wave. All benchmark tests showed excellent agreement with analytic solutions at all drag regimes while conserving mass, momentum, and energy. We additionally demonstrated the robustness of our numerical scheme on two sets of more complex scenarios: Dusty Cold Keplerian Disks, Dusty protoplanetary disks, and Dusty Kelvin-Helmholtz instabilities. Thanks to the dust diffusion model, the code proved to be stable under most parameter choices, the only exception being the weak and intermediate drags ($K=0.01$ and $K=1$) for a dust-to-gas ratio of unity. Future improvements will involve implementing the One Fluid model also to the Meshless-Finite-Mass framework in \texttt{OpenGadget3} \citep{Groth2023} as well as extending the implementation to multiple grain sizes.\\

\begin{acknowledgements}
    TB acknowledges funding by the Deutsche
    Forschungsgemeinschaft (DFG, German Research Foundation) under grant 325594231. GTP, TB, KD, and BE acknowledge support by the Deutsche Forschungsgemeinschaft (DFG, German Research Foundation) under Germany's Excellence Strategy - EXC-2094 - 390783311.  GTP and KD acknowledge support by the COMPLEX project from the European Research Council (ERC) under the European Union's Horizon 2020 research and innovation program grant agreement ERC-2019-AdG 882679. GTP and TB acknowledge funding from the European Union under the European Union's Horizon Europe
    Research and Innovation Programme 101124282 (EARLYBIRD). Views and opinions expressed are, however, those of the
    authors only and do not necessarily reflect those of the European Union or the European Research
    Council. Neither the European Union nor the granting authority can be held responsible for them. MH acknowledges support from the DFG program ``Closing the Loop - Using Synthetic Observations of Simulated Star-forming Regions to Test Observational Properties'' (DFG Project Number: 426714422).
\end{acknowledgements}

\bibliographystyle{aa}
\bibliography{bibliography}

\end{document}